# PARAMETER-AWARE HIGH-FIDELITY MICROSTRUCTURE GENERATION USING STABLE DIFFUSION

Hoang Cuong Phan[a,b], Minh Tien Tran[a], Chihun Lee[a], Hoheok Kim[a], Sehyeok Oh[a], Dong-Kyu Kim[c], Ho Won Lee[a,b,*]

[a] *Materials Data & Analysis Research Division, Korea Institute of Materials Science, Changwon 51508, Republic of Korea*

[b] *Advanced Materials Engineering, University of Science and Technology, Daejeon 34113, Republic of Korea*

[c] *Department of Mechanical Engineering, Konkuk University, Seoul 05029, Republic of Korea*

*Corresponding author

Ho Won Lee

Phone: +82-55-280-3843

Fax: +82-55-280-3637

E-mail: h.lee@kims.re.kr

**Abstract**

Synthesizing realistic microstructure images conditioned on processing parameters is crucial for understanding process-structure relationships in materials design, but it remains challenging due to limited availability of training micrographs and the continuous nature of processing variables. To address these challenges, we present a novel parameter-aware generative modeling approach based on Stable Diffusion 3.5 Large (SD3.5-Large), a state-of-the-art text-to-image diffusion model adapted for microstructure generation. Our method introduces parameter-aware embeddings that encode continuous and categorical variables directly into the model's conditioning, enabling controlled image generation under specified processing conditions and capturing parameter-driven microstructural variations. To overcome data scarcity and computational constraints, we fine-tune only a small fraction of the model's weights using DreamBooth and Low-Rank Adaptation (LoRA), efficiently transferring the pre-trained model to the materials domain with minimal new training data. To validate microstructural realism, we have developed a semantic segmentation model based on a fine-tuned U-Net with a VGG-16 encoder on a limited labeled dataset of 24 experimental micrographs. This approach significantly outperforms previous methods in both accuracy (97.1%) and mean intersection over union (mIoU, 85.7%), ensuring reliable segmentation masks for statistical analyses. Consequently, quantitative analyses using both physical descriptors and spatial statistical functions show strong agreement between synthetic and real microstructures. Particularly, the spatial statistics evaluated via two-point correlation and lineal-path functions yield errors below 2.1% and 0.6%, respectively. These results underscore the novelty and effectiveness of our parameter-aware diffusion approach. To our knowledge, this study marks the first comprehensive adaptation of SD3.5-Large for parameter-aware microstructure generation, providing a scalable and efficient approach applicable to broader scientific domains. These findings highlight the transformative potential of advanced diffusion models to accelerate data-driven materials design.

## 1. Introduction

Understanding the complex relationship between processing parameters, microstructure evolution, and material properties is a fundamental challenge in materials science [1,2]. The process-structure-property paradigm describes how variations in process conditions such as annealing temperature, time, and cooling rate can influence microstructural morphology, phase distribution, and material performance. Microstructure characterization and evolution have traditionally relied on experimental techniques like scanning electron microscopy (SEM), electron backscatter diffraction, alongside computational methods such as crystal plasticity finite element method [3,4]. While these approaches provide valuable insights, they are often time-intensive, data-limited, and computationally expensive. Such constraints hinder rapid exploration of processing scenarios in modern materials development, highlighting a need for more efficient microstructure generation and analysis tools.

In recent years, deep learning-based generative models have emerged as a powerful tool for synthetic microstructure generation, addressing data scarcity by reducing reliance on experiments and simulations. Various methods, including variational autoencoders (VAEs), generative adversarial networks (GANs), and more recently, diffusion models have been explored, but each has inherent drawbacks. VAEs struggle with blurry outputs and loss of fine details [5] due to latent space compression. This limitation reduces their effectiveness in capturing the intricate features necessary for realistic microstructure generation. GANs can generate sharp, detailed microstructure images [6], but suffer from mode collapse [7] and unstable training [8], resulting in limited diversity. Denoising diffusion probabilistic models (DDPMs) [9] provide higher-quality and more diverse generations compared to VAEs and GANs [10–14]. However, DDPMs require large datasets and high computational costs, which can limit their practicality for materials science applications, where data scarcity and computational resources are often a concern [15]. These trade-offs highlight that, despite promising progress, existing generative approaches have not yet fully met the needs of materials science for reliable, high-fidelity microstructure synthesis under data-limited conditions.

Beyond model-class limitations, most microstructure generation studies vary only a few process variables and rarely model their joint effects. For example, recent works consider a single variable (e.g., laser power, cooling regime) without modeling interactions with annealing temperature and time [16,17]. Conditioning is frequently discrete (class labels) rather than continuous, limiting realism when process variables are numeric and multi-dimensional [18,19]. By contrast, work on non-materials datasets embeds continuous variables (e.g., steering angle, chairs) [20,21] but targets a single scalar, leaving multi-dimensional process dynamics unaddressed. General-purpose text-to-image diffusion (e.g., Stable Diffusion) can synthesize microstructure-like images from prompts [22], yet numbers and units in text prompts can be misinterpreted, yielding inaccurate morphologies and disrupted feature layout. Many implementations also train from scratch on small corpora, increasing compute and overfitting risk in data-scarce settings. Validation typically relies on generic vision metrics such as Fréchet Inception Distance (FID) [23], which are not calibrated to microstructural morphology; materials-specific descriptors (e.g., particle-size distributions, two-point correlations) are rarely reported. To our knowledge, few studies pair generation with semantic segmentation to extract morphologies and compute materials-specific statistics end-to-end. Taken together, these observations define three persistent gaps in parameter-aware microstructure generation: (1) joint conditioning on multiple continuous (e.g., temperature, time, magnification) and categorical (e.g., cooling method, microconstituent class) variables; (2) parameter- and data-efficient adaptation, as full fine-tuning or from-scratch training on small, heterogeneous materials datasets risks overfitting and prohibitive computational cost; and (3) quantitative validation beyond visual plausibility, requiring reliable segmentation coupled with physical and spatial descriptors.

We address these gaps by adapting SD3.5-Large [24] to generate microstructures conditioned on processing and acquisition variables. At a high level, we augment the model with a hybrid conditioning scheme that fuses learned embeddings of multiple continuous (temperature, time, magnification as an acquisition parameter) and categorical (cooling method, microconstituent) variables with textual tokens, enabling the denoiser to attend jointly to words, numbers, and categories. To remain data and compute-efficient on small corpora, we combine DreamBooth [25] (data-efficient concept anchoring) with LoRA [26] (low-rank adapters on attention projections) rather than full fine-tuning. We then validate realism by training a lightweight U-Net (VGG-16 encoder [27]) on a small labeled subset and computing physical descriptors and spatial statistics from the predicted masks.

The main contributions of this paper are listed as follows:

**(1) Hybrid conditioning for multiple variables:** we inject multiple continuous and categorical conditions directly into the text-encoder token streams, mitigating numeric misinterpretation in text-only prompts and enabling joint control of process and acquisition variables.

**(2) Parameter-efficient adaptation of SD3.5-Large:** a DreamBooth+LoRA recipe updates only attention-layer adapters and small conditioning modules, retaining strong priors while reducing the number of trainable parameters, which is well-suited to limited micrographs.

**(3) Quantitative validation pipeline:** a VGG-16 U-Net supplies masks for computing materials-specific descriptors to assess fidelity rigorously.

**(4) Comparative study:** we benchmark SD3.5 variants against SDXL, a GAN, and a DDPM baseline adapted to microstructure generation demonstrate improvements in fidelity, statistics, and parameter controllability.

**(5)** We also demonstrate portability on the Aachen-Heerlen dataset without architectural changes.

## 2. Materials and methods

### 2.1. Two-stage parameter-aware generation and validation

We propose a modular, two-stage framework for parameter-aware microstructure generation and validation (Fig. 1). Stage 1 adapts SD3.5-Large using parameter-efficient updates (DreamBooth + LoRA) and structured conditions extracted from the ultra-high carbon steel (UHCS) dataset [28]. Conditions combine textual descriptions with continuous and categorical variables. Stage 2 applies a VGG-16 U-Net segmenter to real and generated micrographs; the resulting masks support quantitative comparison via physical descriptors (e.g., particle size, area fraction) and spatial statistics (e.g., two-point and lineal-path functions). The framework transfers across datasets with different condition schemas. We adopt a modular pipeline, as exemplified by Diffusers [29].

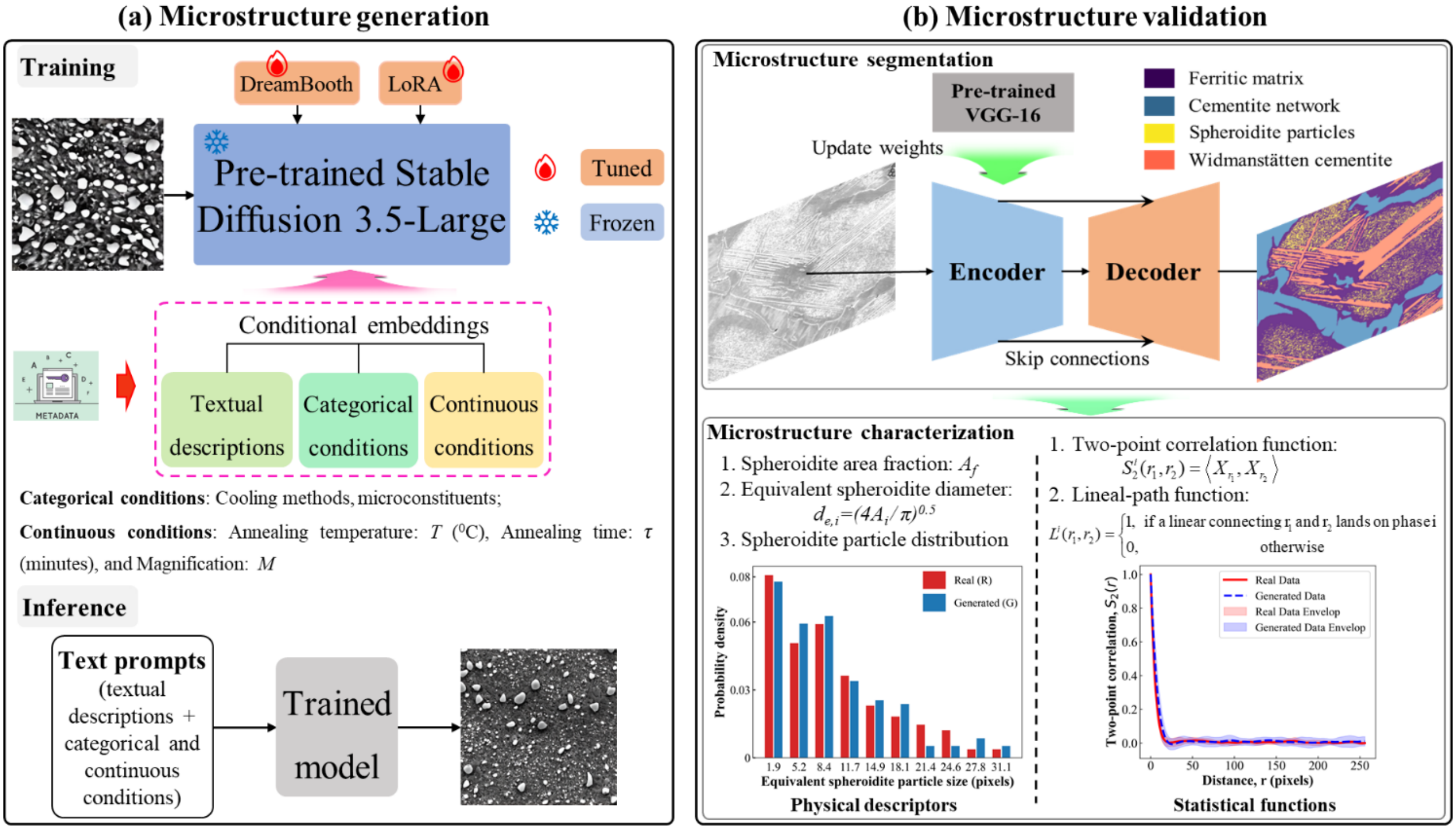

Fig. 1. Two-stage workflow for parameter-aware microstructure generation and validation. (a) Stage 1: Fine-tune SD3.5-Large using textual descriptions with continuous and categorical conditions. (b) Stage 2: Perform semantic segmentation of real and generated images to identify microconstituents. Quantitatively characterize microstructures using physical descriptors and spatial statistical measures.

### 2.2. Ultra-high carbon steel dataset

The dataset used for microstructure generation is sourced from the open literature on UHCS. This dataset consists of 961 SEM micrographs that were taken from different annealing conditions, with 598 images containing detailed metadata on heat treatment conditions, including annealing temperature, time, cooling methods, and magnifications. A visual representation of this dataset is shown in Fig. 2. Most of the micrographs are spheroidite, pearlite, and network morphology, while a smaller subset includes images with mixed two primary microconstituents, such as Widmanstätten cementite, pearlite containing spheroidized cementite, and martensite (Fig. A1). Each micrograph (645 × 481 pixels) is paired with a structured text prompt designed for model training. These prompts combine detailed microstructural descriptions (e.g., spheroidite with cementite network) and numerical parameters (e.g., 800 °C, 480 minutes, 4910× magnification) extracted from an SQLite database [30]. Images are preprocessed to preserve aspect ratio, followed by center-cropping to 512 × 512 pixels, and normalizing to [-1,1] range, to align with SD3.5-Large's input requirements. We used 521 image-text pairs for training and 77 for validation. The validation set is further split into 23 'seen' cases and 54 'unseen' cases (new combinations, notably all 180-minute annealing time cases) to test generalization (see Table A1).

For the segmentation task, a subset of the UHCS dataset [31], consisting of 24 manually labeled images, is utilized. This subset contains only spheroidite (Fig. 2(*e*)) and spheroidite+Widmanstätten (Fig. 2(*f*)) micrographs. These images contain pixel-wise annotations for different microconstituents, enabling precise segmentation model training.

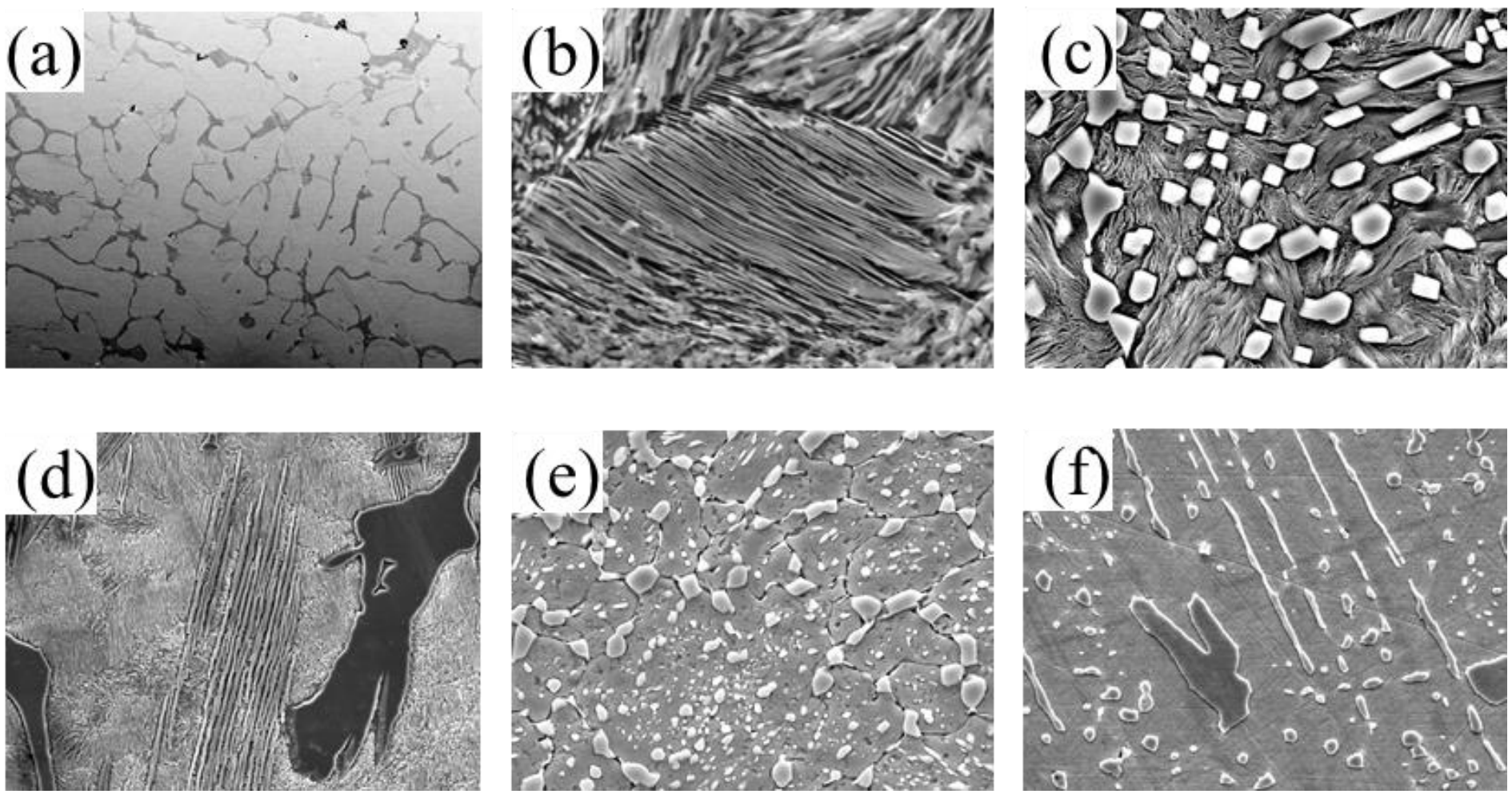

Fig. 2. Representative microstructures in the UHCS dataset (primary constituents shown): (a) proeutectoid cementite network (N), (b) pearlite (P), (c) pearlite + spheroidite (PS), (d) pearlite + Widmanstätten (PW), (e) spheroidite particles (S), and (f) spheroidite + Widmanstätten (SW) [28].

### 2.3. Parameter-efficient fine-tuning of SD3.5-Large with parameter-aware conditioning

Prior microstructure generators commonly discretize experimental variables or control only a single continuous label, limiting controllability across realistic processing windows. Adapting large diffusion models to data-scarce, domain-specific problems compounds this challenge: although rectified-flow formulations enable high-resolution synthesis, naively fine-tuning all weights is compute-intensive and prone to overfitting or catastrophic forgetting. We therefore adapt SD3.5-Large with parameter-efficient updates and a token-level parameter-aware conditioning scheme that injects continuous and categorical parameters directly into the text encoders. The model then cross-attends jointly to words, numbers, and categories at every denoising step, while the base weights remain frozen.

2.3.1. DreamBooth and LoRA for efficient adaptation

We personalize SD3.5-Large to the microstructure domain with DreamBooth [25], then make adaptation feasible with LoRA [26]. DreamBooth anchors domain concepts from a smaller number of exemplars while preserving generality via prior preservation. To avoid updating the full network, we insert low-rank adapters in attention projections and update only those, together with the small conditioning modules introduced below. For a linear map $W_0 \in R^{d\times k}$, LoRA learns a low-rank $r$ update:

$$\Delta W = BA, \quad W = W_0 + \Delta W \tag{1}$$

with low-rank matrices $B \in R^{d\times r}$, $A \in R^{r\times k}$, $r \ll \min(d,k)$. This retains the strong pre-trained priors of SD3.5-Large while keeping the trainable footprint small. The combined DreamBooth + LoRA yields an adapted model, but still relies on textual prompts lacking parameter-awareness.

2.3.2. Hybrid conditioning over continuous and categorical parameters

To enable multi-parameter control, we extend prompts with learned embeddings, replacing placeholders in CLIP-L/G sequences as shown in Fig. 3. SD3.5-Large uses three frozen text encoders for handling the text conditioning, CLIP-L/14 [32] (local semantics), OpenCLIP-G/14 [33] (global semantics), and T5-XXL [34] (long-form context). We condition at the token level in the two CLIP streams and then fuse all encoders.

For multiple continuous variables, let $u = [u_1, \dots, u_N] \in R^N$ collect the continuous conditions supplied by a dataset schema (e.g., for UHCS: temperature $T$ [°C], time $\tau$ [min], magnification $M$; for Aachen-Heerlen (A-H) [35]: temperature $T$ [°C], distance $d$ [mm]). To stabilize training across different ranges, each component uses a robust, dataset-aware normalization (with an optional log transform for variables with wide dynamic range, as specified by the schema):

$$\tilde{u}_i = \frac{u_i - median(u_i)}{Q_3(u_i) - Q_1(u_i)}, \quad \tilde{u} = [\tilde{u}_1, \dots, \tilde{u}_N], \tag{2}$$

and for heavy-tailed variables we additionally clip values outside [$Q_1$ - 1.5IQR, $Q_3$ + 1.5IQR], where IQR = $Q_3$ - $Q_1$. For each CLIP encoder $j \in \{L, G\}$ with hidden width $d_j$, a small MLP $f_j$: $R^N \rightarrow R^{d_j}$ maps $\tilde{u}$ to a token-sized continuous embedding

$$e_{num}^{(j)} = W_2^{(j)}.\sigma\left(W_1^{(j)}.\tilde{u} + b_1^{(j)}\right) + b_2^{(j)} \in R^{d_j}, \tag{3}$$

with SiLU nonlinearity $\sigma$. This vector is inserted at each continuous placeholder position (e.g., *<temp>*, *<time>*, *<mag>*) in the corresponding CLIP token sequence.

For each categorical key (e.g., cooling method, microconstituent; or dataset-specific categories in Aachen-Heerlen) we add a dedicated placeholder token and a per-encoder embedding table $E_{cat}^{(j)} \in R^{K\times d_j}$ with K categories. The selected category vector

$$e_{cat}^{(j)} \in R^{d_j} \tag{4}$$

replaces the embedding at the categorical placeholder positions (e.g., *<cool>*, *<micro>*). Placeholders are added to the prompt automatically when absent to ensure consistent positions.

Token-level replacement: denote $H^{(j)} \in R^{n_j \times d_j}$ the CLIP token sequence (hidden states) before replacement, and by $I_{num}^{(j)}, I_{cat}^{(j)} \subset \{1, \dots, n_j\}$ the sets of indices of continuous and categorical placeholders. We form the modified sequences in one step:

$$H_{cond}{}^{(j)} = Replace\left(H^{(j)};\ I_{num}^{(j)} \ \rightarrow\ e_{num}^{(j)},\ I_{cat}^{(j)} \ \rightarrow\ e_{cat}^{(j)}\right),\ j \in \{L, G\} \tag{5}$$

This ensures that numbers and categories participate in cross-attention wherever semantics are formed. T5-XXL is appended unchanged (see Fig. 3).

2.3.3. Conditioning fusion

From the modified CLIP streams, we compute two conditionings used by the denoiser, adapting the SD3.5-Large fusion while incorporating our modifications to the CLIP text encoders. The pooled outputs from the modified CLIP-L and CLIP-G encoders (size 768 and 1280, respectively) are concatenated along the channel axis to obtain the vector conditioning:

$$c_{vec} = Concat(p_L; p_G)\ \in\ R^{2048} \tag{6}$$

Token sequences are concatenated along the channel axis, with modified CLIP encoders zero-padded to match the T5-XXL width:

$$c_{ctxt} = Concat\left(Pad(H_{cond}^{(L)}; H_{cond}^{(G)}\right), H_{T5})\ \in\ R^{154 \times 4096} \tag{7}$$

The diffusion transformer conditions on $c$ = $(c_{vec}, c_{ctxt})$ at each denoising step, so words, continuous tokens, and categorical tokens are all visible to attention.

2.3.4. Training objective and sampling

We adopt the rectified-flow formulation from Stable Diffusion 3 [24]. Let $y_t$ follows the ordinary differential equation (ODE):

$$\frac{dy_t}{dt} = v_\theta(y_t, t) \tag{8}$$

With a forward process that transports a real micrograph by $x_0 \sim p_0$ to noise $\epsilon \sim N(0, I)$, and a continuous time $t \in [0,1]$ along a straight path:

$$z_t = (1 - t)x_0 + t\epsilon \tag{9}$$

Training minimizes a conditional flow-matching (CFM) objective that is equivalent to a weighted noise-prediction loss:

$$L_{CFM} = E_{t, p_t(Z|\epsilon), p(\epsilon)} \| v_\theta(z, t, c) - u_t(z|\epsilon) \|_2^2 \tag{10}$$

Where $u_t(\cdot)$ is the rectified flow, and c denotes the conditioning we defined in Section 2.3.3. This follows the simulation-free conditional flow-matching framework and its rectified-flow specialization in SD3-class model.

2.3.5. Selective updates and scope

Trainable parameters are restricted to (i) LoRA adapters on the attention projections (Q, K, V, and the output projection (O)) in multimodal diffusion transformers (MM-DiT) and in both CLIP text encoders (CLIP-L/14 and CLIP-G/14), and (ii) the conditioning modules: per-encoder MLPs for continuous variables and per-encoder embedding tables for categorical variables. All other weights remain frozen. A component-wise breakdown of total and trainable parameters for SD3.5-Large is provided in Table B3.

To evaluate the approach, we compare against Stable Diffusion variants and classic generative baselines. We fine-tune Stable Diffusion 3.5 Medium (SD3.5-Medium) using the same adaptation recipe as SD3.5-Large. We also assess Stable Diffusion XL (SDXL) [36] under three adapter strategies: (i) LoRA alone, (ii) DreamBooth and LoRA, and (iii) DreamBooth combined with Weight-Decomposed Low-Rank Adaptation (DoRA) [37]. In addition, we benchmark StyleGAN3 [38] and DDPM [9,39,40] baselines adapted to microstructure synthesis under matched data and computational conditions.

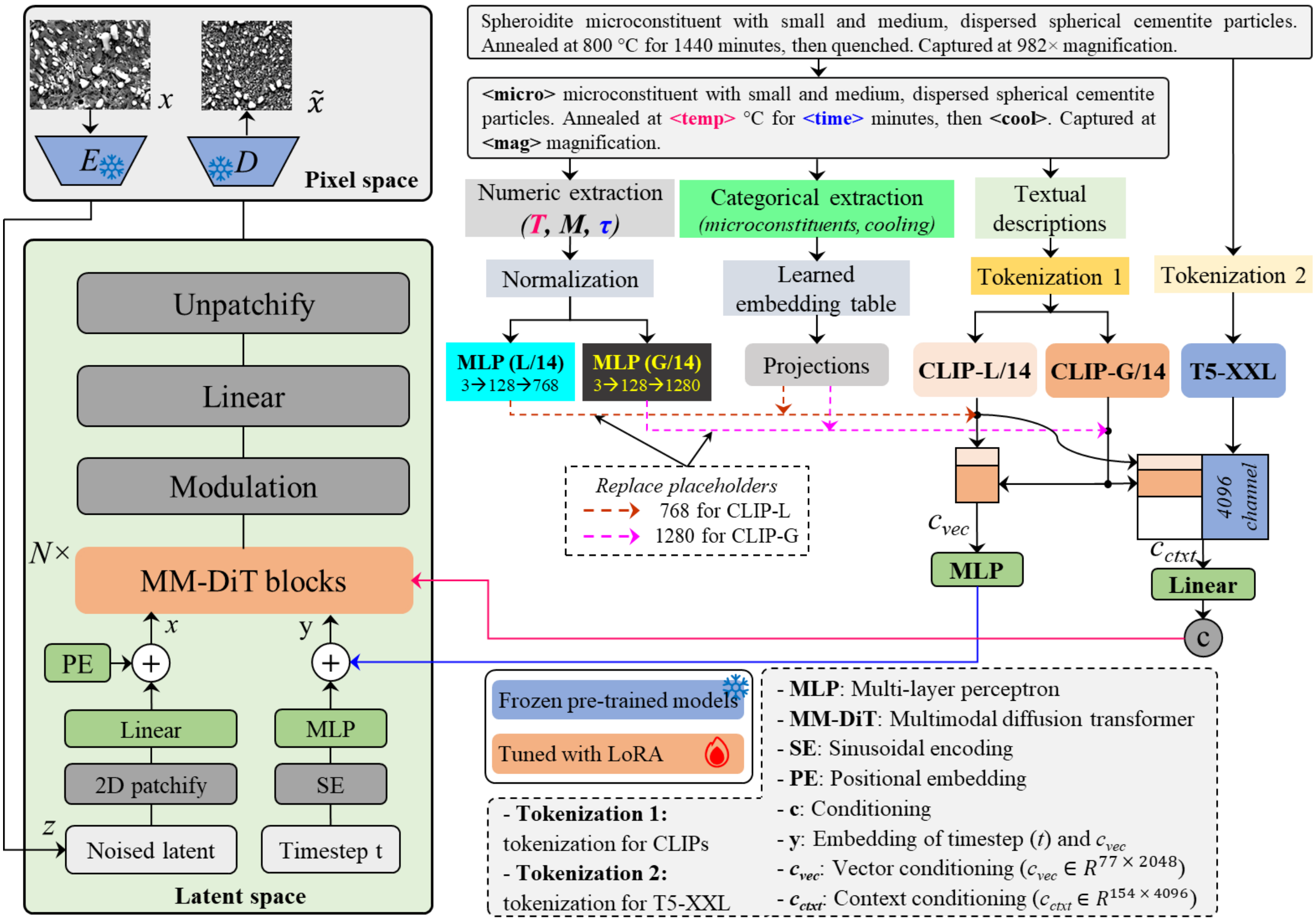

Fig. 3. Parameter-aware token conditioning and parameter-efficient adaptation. Schema-defined continuous and categorical placeholders in the CLIP-L/14 and OpenCLIP-G/14 token streams are replaced with learned embeddings derived from dataset metadata; T5-XXL provides long-range text context unchanged. The denoiser attends jointly to words, numbers, and categories, while LoRA adapters (red) enable parameter-efficient fine-tuning; the VAE remains frozen.

### 2.4. Semantic segmentation for quantitative microstructure characterization

Accurate microconstituent–resolved segmentation is essential for validating generated microstructures, particularly in complex, anisotropic, multi-phase systems such as those in the UHCS dataset. Without accurate segmentation of microconstituents, we cannot quantitatively compare real and generated microstructures beyond superficial metrics. Thus, a robust segmentation model is essential to our validation strategy. Recent machine learning advances in unsupervised [41], semi-supervised [42], and supervised [43–45] segmentation have improved accuracy by learning complex patterns; however, their integration with quantitative microstructure characterization remains limited by scarce labeled datasets [46,47].

Applying semantic segmentation to microstructure images presents several challenges. These include unbalanced class distributions, high variability in shape, size, and textures among microconstituents, high-resolution images, and the absence of prior structural information [44]. Additionally, the scarcity of large-scale datasets in materials science limits the ability to pre-train models on extensive datasets and fine-tune them on smaller, domain-specific sets. Furthermore, generating accurate ground truth labels for intricate microstructural images requires expertise from materials scientists, making it both time-consuming and costly.

To address these issues, we propose a data-efficient semantic segmentation framework that performs reliably with limited annotation. The model automates microconstituent identification and enables precise calculation of physical descriptors and statistical functions used later for validation.

We adopt a U-Net with a pre-trained VGG-16 encoder (Fig. 4). All encoder layers are fine-tuned, and the decoder is trained from scratch. The encoder retains Conv followed by ReLU activations. Decoder blocks use Conv, GroupNorm, and LeakyReLU. Training uses AdamW with separate learning rates for the encoder and the decoder. Data augmentation includes horizontal flips, random crops, brightness and contrast jitter, and rotations in the range 0 to 45$^{o}$. We compare standard losses (categorical cross-entropy, Dice, Focal, and Jaccard) and select Jaccard based on class-wise performance. Exact learning rates, batch size, weight decay, regularization settings, and augmentation ranges are provided in Appendix C.

To emphasize informative channels under class imbalance, we insert one Efficient Channel Attention (ECA) [48] module after the final convolution of each encoder stage (Conv1 through Conv5), as shown in Fig. 4(b). Skip connections originate after ECA, so the decoder receives channel-reweighted features. As illustrated in Fig. 4(c), ECA computes channel weights by global average pooling (GAP) followed by a lightweight one-dimensional convolution and a sigmoid, and applies multiplicative reweighting to the feature maps. We use a fixed kernel size $k = 5$ in this work.

This design yields a segmentation model that is reproducible, robust under scarce labels and class imbalance, and directly supports quantitative validation of the generated microstructures.

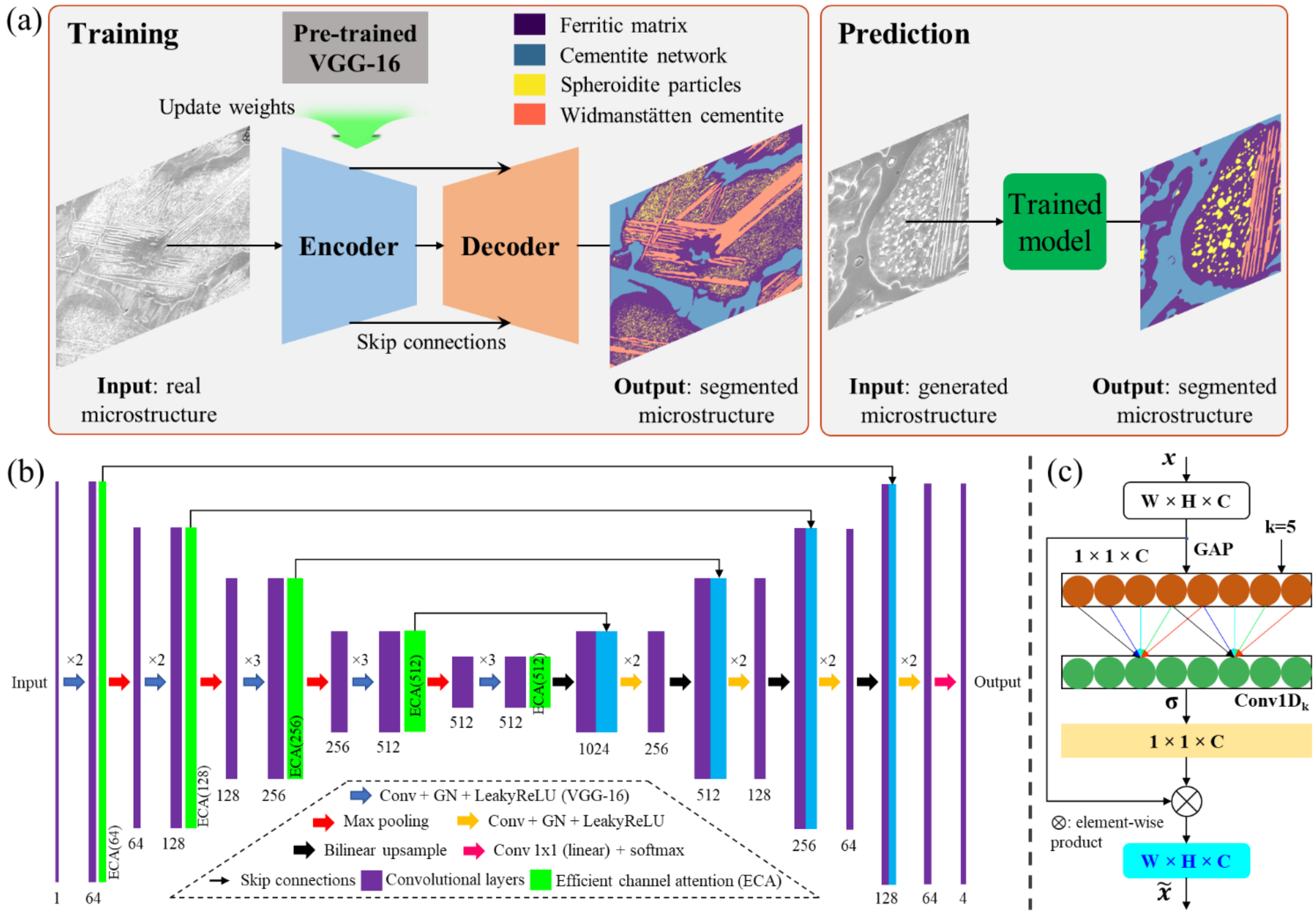

Fig. 4. Semantic segmentation framework for microstructure validation. (a) U-Net with a pre-trained VGG-16 encoder is fine-tuned on the labeled UHCS micrographs and then applied to real/generated images to predict four microconstituent classes. (b) Detailed architecture with efficient channel attention (ECA, green) integrated after each encoder convolutional block (purple), enabling channel-wise refinement without dimensionality reduction, alongside group normalization, LeakyReLU activations, max pooling (red arrows), bilinear upsampling (green/yellow arrows), and skip connections (black) for robust feature fusion despite class imbalance and data scarcity. (c) Zoomed ECA module: post-conv features (W×H×C) undergo global average pooling (GAP) to 1×1×C, followed by 1D convolution with kernel k=5, sigmoid activation, and element-wise multiplication (×) back to original features for efficient cross-channel interaction [48].

### 2.5. Physical and statistical metrics for microstructure evaluation

Evaluation metrics in image recognition tasks primarily assess visual similarity based on human perception, they may not fully capture the intricacies of microstructures. Therefore, we employ both physical descriptors and statistical functions for microstructure characterization.

Physical descriptors in this study include deterministic (area fraction of spheroidite, $A_f$) and statistical (equivalent spheroidite diameter $d_{e,i}=(4A_i/\pi)^{0.5}$ and spheroidite particle distributions). For statistical functions, spatial correlation functions are used to analyze the spatial and morphological characteristics of generated microstructures, as detailed in [15]. The two-point correlation ($S_2(r)$) measures the probability of finding two random points separated by a vector $r$ within a specific material phase [49,50]. The lineal-path function [50] ( $L(r)$ ) evaluates connectivity between clusters within a material phase, determining whether a direct path between two points lies entirely within the phase. Differences between real and generated microstructure images are quantified as $\varepsilon_c = A_1/A_2$ (%). Here, $A_1$ represents the area between their correlation functions, while $A_2$ is the area under the real microstructure's correlation function.

## 3. Results and discussion

### 3.1. Fidelity and diversity of synthetic microstructures

#### 3.1.1. Cross-family benchmarking with parameter-only conditioning

To assess how well different generative paradigms capture parameter-aware microstructures, we benchmark three models under identical parameter-only conditioning (i.e., continuous variables such as temperature, time, and magnification, plus categorical ones like cooling method and microconstituent, without additional textual prompts): StyleGAN3 and DDPM, both trained from scratch, and SD3.5-Large fine-tuned with DreamBooth and LoRA. Detailed descriptions and discussion of metrics are provided in Sections A2 and B1.

As summarized in Table 1, SD3.5-Large fine-tuned with DreamBooth + LoRA delivers the best overall fidelity and coverage (lower CMMD/LPIPS; higher Precision/Recall) while reducing training time by roughly 9× faster than from-scratch baselines in our setup. StyleGAN3 is the fastest at inference but exhibits lower Recall, indicating narrower support of the real distribution. DDPM improves Recall over StyleGAN3 but remains costly in both training and sampling. In contrast, parameter-efficient fine-tuning of a large diffusion prior offers substantially better quality and competitive sampling speed. The hyperparameters for these models are presented in Tables B1, B2.

Table 1. Quantitative comparison under parameter-only conditioning on the UHCS validation set. Lower is better for CMMD and LPIPS; higher is better for SSIM, Precision, Recall. Training time measured to validation plateau on a single A100-80 GB; sampling time on a single RTX A6000.

| Model | CMMD (↓) | SSIM (↑) | LPIPS (↓) | Precision (↑) | Recall (↑) | Training time (h) (↓) | Sampling time (sec/image) (↓) |
|---|---|---|---|---|---|---|---|
| StyleGAN3 | 1.555 | 0.125 | 0.585 | 0.547 | 0.171 | 63.6 | **0.2** |
| DDPM | 0.874 | 0.253 | 0.543 | 0.611 | 0.408 | 64.4 | 20.6 |
| SD3.5-Large + DreamBooth and LoRA | **0.700** | **0.607** | **0.473** | **0.701** | **0.500** | **7.4** | 13 |

Fig. 5 presents real micrographs alongside images generated by StyleGAN3, DDPM, and SD3.5-Large under parameter-only conditioning for two seen (top rows) and two unseen (bottom rows) process conditions. Labels encode the conditioning variables; for example, '800C-180M-982×-Q-S' denotes continuous conditions of 800 $^{o}$C for 180 minutes, at 982× magnification; and categorical conditions of quenching (Q) and spheroidite (S). SD3.5-Large yields the highest visual fidelity, whereas StyleGAN3 tends to produce softer, less distinct textures. For all models, exact replication of microstructure shape, morphology, position, and orientation remains challenging, particularly for minority classes such as Widmanstätten. Additional quantitative and qualitative results (Fig. B9 and Table 6) corroborate the overall advantage of SD3.5-Large while revealing residual artifacts and occasional mis-generated elements, motivating the enhanced conditioning introduced next.

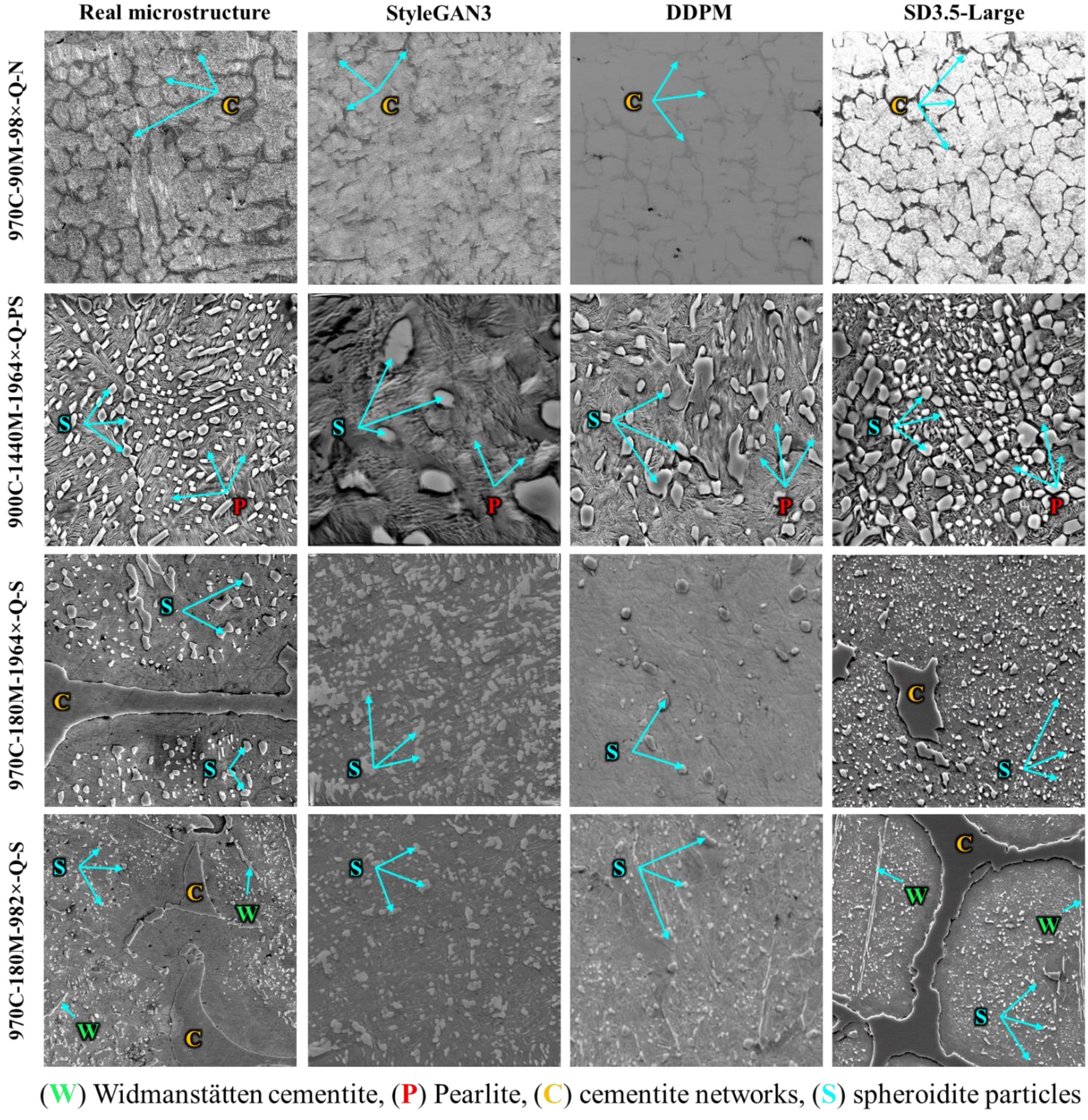

Fig. 5. Parameter-only qualitative comparison across two seen and two unseen conditions. Rows (conditions): (1) 970°C, 90 min, 98×, quenched, network; (2) 900°C, 1440 min, 1964×, quenched, pearlite + spheroidite; (3) 970°C, 180 min, 1964×, quenched, spheroidite; (4) 970°C, 180 min, 982×, quenched, spheroidite. Real micrographs on left followed by generations from StyleGAN3, DDPM, and SD3.5-Large. Arrows highlight key microconstituents.

3.1.2. Effect of text prompt types

Table 2. Effect of prompt types on SD3.5-Large generation performance.

| Types of text prompts | CMMD (↓) | CLIP Score (↑) | SSIM (↑) | LPIPS (↓) | Precision (↑) | Recall (↑) |
|---|---|---|---|---|---|---|
| Parameter-only | 0.7 | - | 0.607 | 0.473 | 0.710 | 0.5 |
| Text-only | 0.591 | 0.320 | 0.649 | 0.440 | 0.868 | 0.711 |
| **Full-text** | **0.538** | **0.321** | **0.654** | **0.429** | **0.921** | **0.789** |

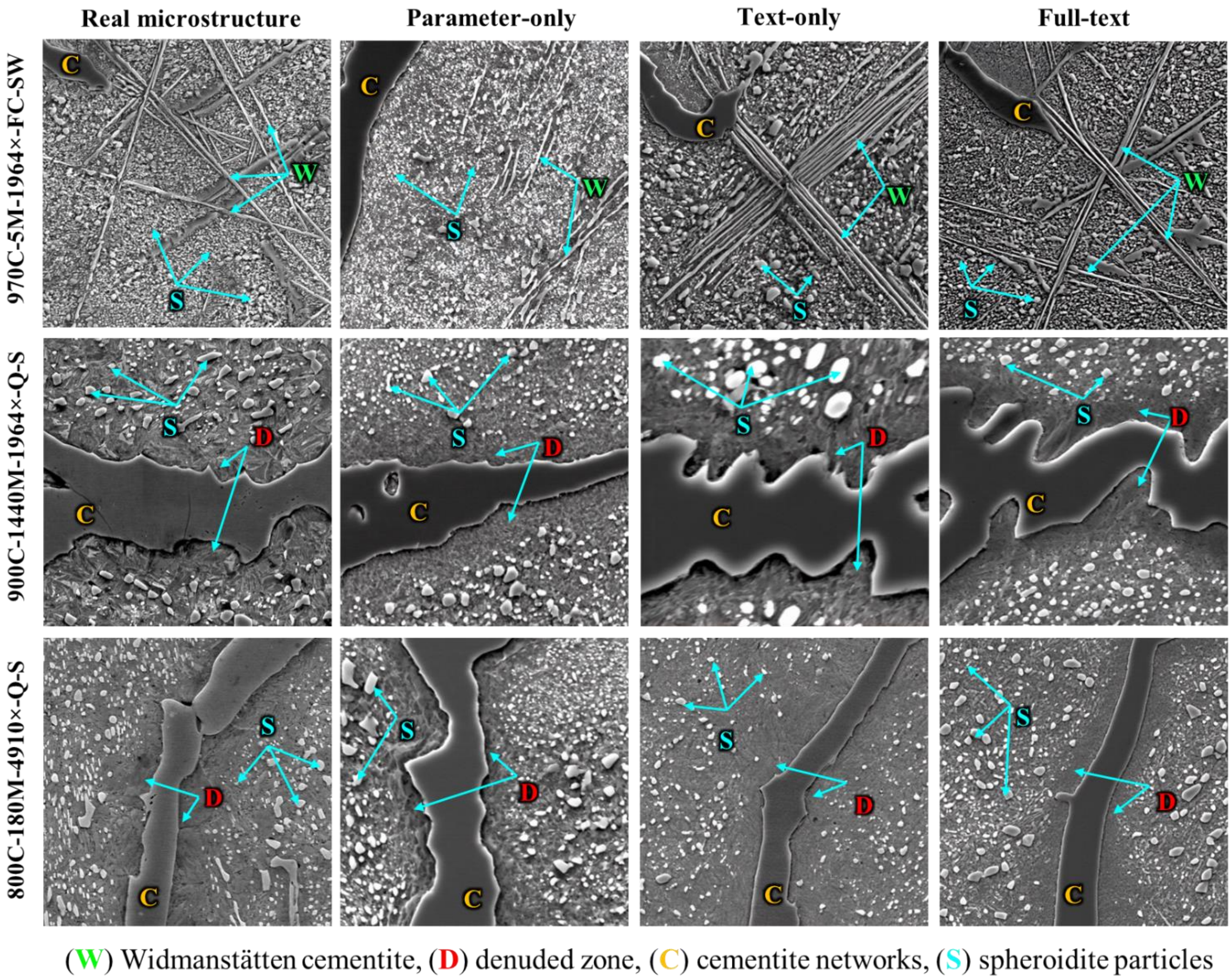

Fig. 6. Impact of prompt types on synthetic microstructure fidelity. Comparison of real microstructures with SD3.5-Large-generated images conditioned on parameters-only, text-only, and full-text prompts. Only full-text prompts enabled by parameter-aware conditioning facilitate accurate generation of key microstructural features, including Widmanstätten cementite, denuded zones, cementite networks, and spheroidite particles.

We evaluated SD3.5-Large under three prompt regimes: parameter-only (continuous and categorical variables, no text), text-only (descriptive phrases, no variables), and full-text (textual descriptions + continuous and categorical conditions). Full-text produced the most faithful microstructures, best matching experimental images (Table 2 and Fig. 6). Text-only preserved overall morphology but lacked quantitative control of feature sizes, orientations, and phase fractions, yielding systematic deviations. Parameter-only consistently underperformed, with shape and distribution artifacts; these trends are corroborated by the evaluation metrics and by t-SNE projections (Fig. B9). Mechanistically, natural-language tokens provide semantic grounding and compositional priors, whereas explicit process parameters supply magnitudes; only their combination delivers both. To further illustrate the effect of parameter-aware conditioning, Fig. B1 visualizes additional examples comparing generations with versus without embeddings across magnifications. Accordingly, we adopt full-text prompts for the remainder of the study. Our choice of SD3.5-Large over SDXL and SD3.5-Medium is corroborated by the full-text ablation in Table B4 and the qualitative comparison in Fig. B2, which together show stronger structural fidelity for SD3.5-Large.

### 3.2. Validation of synthetic microstructures

#### 3.2.1. Microstructure segmentation accuracy

We evaluate semantic segmentation on the UHCS subset [31] using a VGG-16 U-Net. Table 3 compares our best model with prior methods. Our configuration with Jaccard loss achieves 85.7% mIoU and 97.1% pixel accuracy, surpassing PixelNet [31] and U-Net++ [44]. Table 4 shows that Jaccard loss yields the strongest performance among the losses considered (Fig. C1 and Table C1). Qualitative results in Fig. 7 illustrate close agreement between predictions and ground truth. Residual errors are confined to the finest scales, particularly along thin Widmanstätten cementite laths.

To analyze robustness under class imbalance and limited labels, we report per-class metrics in Table 5 and visualize errors in Fig. 8. Majority microconstituents Class 0 (ferritic matrix) and Class 1 (cementite network) exhibit precision and recall above 96%, with very low FP/FN. Class 3 (Widmanstätten cementite laths) remains most challenging: precision and recall drop to about 79-80%, and nearly 92% for Boundary-F1 (BF1), reflecting boundary-localization errors on thin laths. The overlays in Fig. 8 confirm that FP/FN concentrate near microconstituent interfaces and along slender Widmanstätten features.

These outcomes stem from three factors. First, severe class imbalance, as Widmanstätten cementite accounts for only approximately 3% of pixels [31,44]. Second, annotation difficulty, identifying thin laths requires expert judgment, and they can be mistaken for grain-boundary cementite [31]. Third, few labeled images (24 in the UHCS subset), which limits learning of rare fine-scale features. A preliminary ablation indicates that adding an ECA module increases Class 3 from 64.7% to 67.3%, suggesting channel attention helps mitigate minority-class errors.

Despite these challenges, the model reliably delineates the major microconstituents and provides dependable masks for downstream validation. In the next section, we use these masks to compute morphological and spatial descriptors (Fig. 9) for quantitative assessment of the generated images.

Table 3. Comparison of segmentation performance with prior methods on the UHCS subset.

| Model | mIoU (%) | Accuracy (%) |
|---|---|---|
| PixelNet [44] | 70.79 | 90.77 |
| U-Net++ (Dice) [44] | 76.46 | 92.51 |
| AMVE 3-best (K1) [44] | 76.71 | 92.49 |
| PixelNet (CCE) [31] | 75.40 | 92.60 |
| PixelNet (Focal) [31] | 62.60 | 86.50 |
| **VGG-16 encoder U-Net (this work)** | **85.70**±1.41 | **97.10**±0.82 |

*AMVE: Artificial MultiView Ensemble; K1: kernel size=1; categorical cross-entropy (CCE)*

Table 4. Semantic segmentation performance across different loss functions averaged over validation images (%).

| Loss | Accuracy | IoU 0 | IoU 1 | IoU 2 | IoU 3 | mIoU |
|---|---|---|---|---|---|---|
| CCE | 97.09±0.77 | 96.26±1.15 | 93.90±2.56 | 84.71±1.90 | 66.59±6.87 | 85.37±1.21 |
| Dice | 96.99±0.83 | 96.13±1.20 | 93.5±2.58 | 84.3±2.04 | 65.75±6.94 | 84.92±1.77 |
| Focal | 96.74±0.79 | 96.29±1.17 | **93.93**±2.22 | 85.31±2.18 | 65.75±6.68 | 85.30±1.66 |
| **Jaccard** | **97.10**±0.82 | **96.34**±1.15 | 93.76±2.35 | **85.40**±1.72 | **67.30**±6.04 | **85.70**±1.41 |

**Class** *0: Ferritic matrix; 1: Cementite network; 2: Spheroidite particles; 3: Widmanstätten cementite*

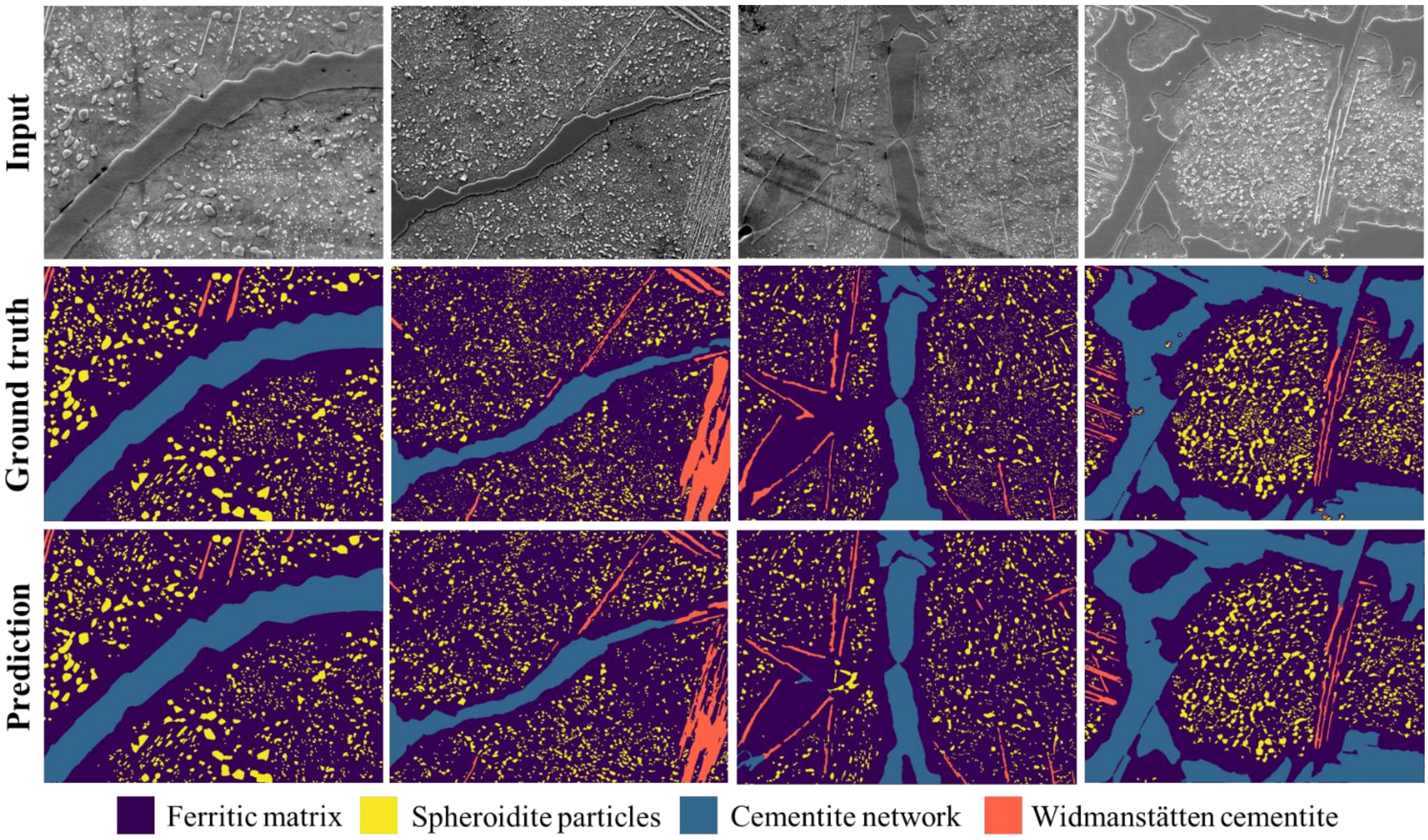

Fig. 7. Qualitative results of our proposed semantic segmentation model on the test set: Input micrographs, ground truth annotations, and predicted segmentation masks.

Table 5. Per-class error metrics (%) using Jaccard loss: Precision, Recall, F1-Score, False Positive (FP), False Negative (FN), and Boundary F1 (BF1). Lower Precision/Recall and BF1 for class 3 indicate boundary-centric errors on thin laths.

| Class | Precision | Recall | F1-Score | FP | FN | BF1 |
|---|---|---|---|---|---|---|
| 0 | 97.67±1.08 | 98.52±0.37 | 98.09±0.63 | 7.09±2.39 | 1.48±0.31 | 98.53±0.44 |
| 1 | 97.97±0.38 | 96.59±2.55 | 96.90±1.46 | 0.30±0.07 | 4.15±3.07 | 93.06±4.46 |
| 2 | 93.53±1.58 | 89.90±2.24 | 91.68±1.28 | 0.72±0.20 | 10.09±2.24 | 98.55±0.24 |
| 3 | 78.70±6.89 | 80.02±11.00 | 79.72±5.19 | 0.32±0.16 | 19.23±11.00 | 91.70±2.96 |
| **Overall** | 91.75±1.32 | 90.85±2.43 | 91.06±1.01 | 2.11±2.88 | 8.74±6.81 | 95.46±4.11 |

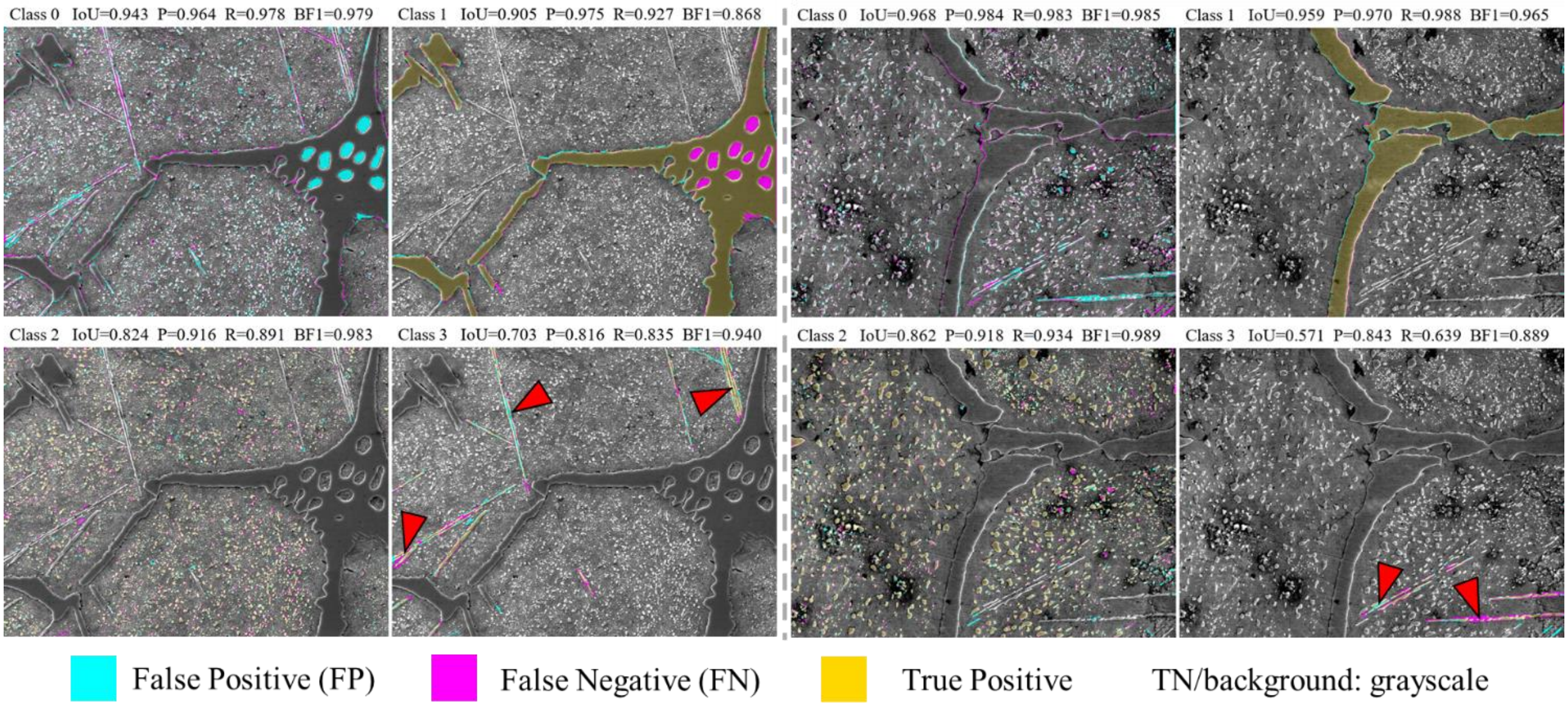

Fig. 8. Per-class error overlays on two validation micrographs. For each micrograph (left and right group), rows correspond to classes: Class 0, ferritic matrix; Class 1, cementite network; Class 2: spheroidite particles; Class 3, Widmanstätten cementite. Colored overlays indicate true positives (yellow), false positives (cyan), and false negatives (magenta); true negatives and background remain grayscale. Panel headers report IoU, Precision (P), Recall (R), and Boundary-F1 (BF1) for the displayed class. Errors cluster at phase boundaries, especially along thin Widmanstätten laths, which appear as magenta FN streaks and occasionally cyan FP at edges. Ferritic matrix and network cementite show crisp boundaries with low FP. Red triangles mark representative failures where laths are so thin that the model confuses them with neighboring microconstituent classes.

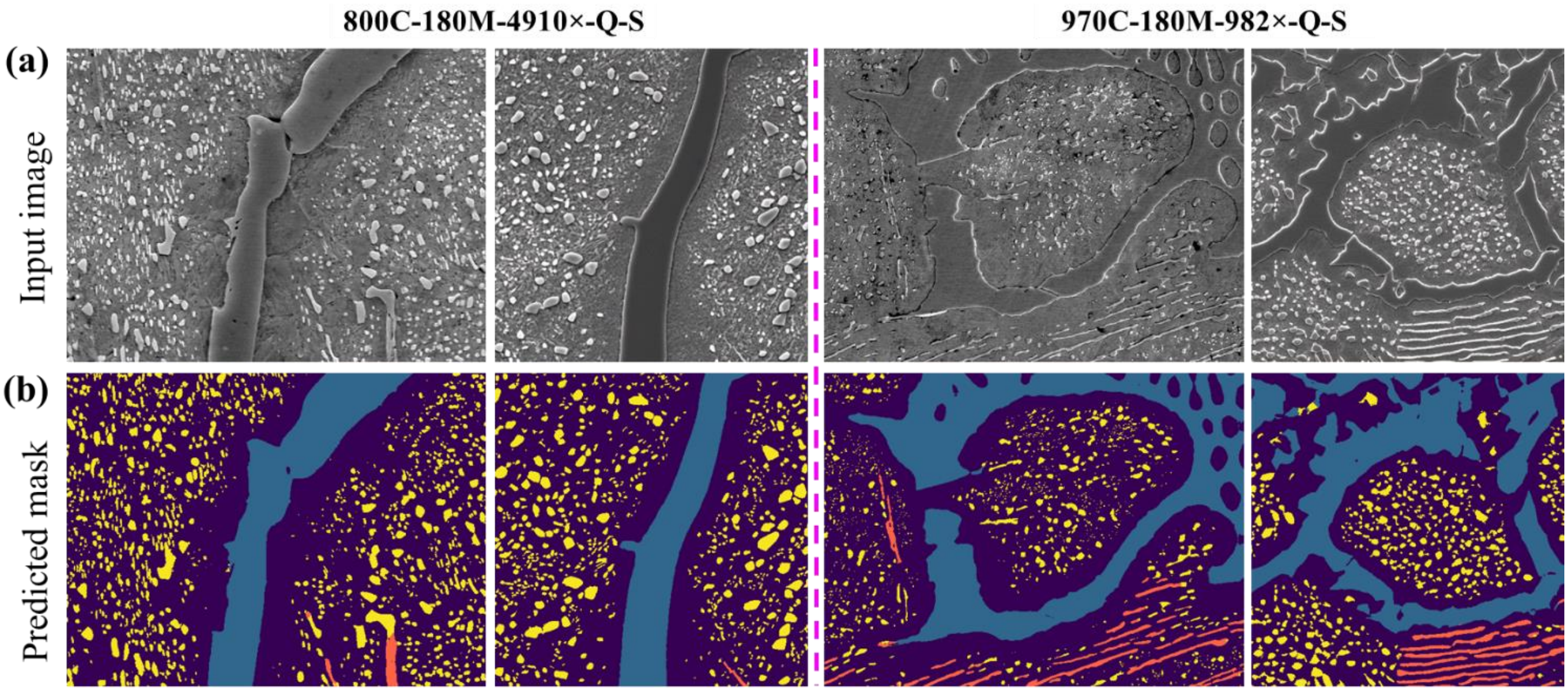

Fig. 9. Testing application of the trained semantic segmentation model on unseen real and generated microstructures under two representative conditions. (a) Input microstructure images (left: real, right: generated), (b) corresponding predicted segmentation masks showing four distinct microconstituents.

### 3.2.2. Microstructure characterization

#### *3.2.2.1. Physical descriptors*

Having validated the segmentation model, we apply it to characterize the microstructures generated by our generative model. Focusing on spheroidite particles, in line with previous studies [31,51], we assess how accurately the model generates key microstructural features under varying processing conditions. The analysis begins with a qualitative comparison to assess visual fidelity and diversity, followed by a quantitative evaluation of spheroidite particle size distributions, mean particle sizes, and area fractions in both real and synthetic micrographs. See Table B5 for detailed fidelity evaluations using additional physical descriptors, including aspect ratios, interfacial areas, and mean nearest neighbor distances [1], which further confirm microstructural realism. Since mean particle size can be skewed by outliers, we report it together with the full size distribution to better capture coarsening trends and to highlight any deviations between real and synthetic images. These comparisons span both seen processing conditions (present in the training dataset) and unseen conditions (interpolated inputs) to test the model's generalization. However, due to lack of real micrographs for unseen annealing times and cooling methods, quantitative metrics for those cases are reported only for seen conditions, as detailed in Sections B2.3(c) and B2.3(d), and visualized in Figs. B7, B8. The following subsections examine the effects of annealing temperature and imaging magnification on the spheroidite microconstituent.

*a. Effect of temperature*

Fig. 10 compares real and generated micrographs across temperatures within the training regime (seen processing cases). Our parameter-aware Stable Diffusion model achieves remarkable visual and statistical fidelity. Generated images closely resemble micrographs in morphology and faithfully capture the spatial distribution and coarsening behavior of spheroidite particles (Fig. 10(a) and Fig. B3). Particle size histograms confirm replication of the full size range, with only minor discrepancies in the largest particle frequencies that do not affect overall accuracy. Quantitatively, the model robustly encodes temperature effects, with spheroidite diameter increasing from 800 $^{o}$C to 970 $^{o}$C, in agreement with an established coarsening mechanism [51]. As shown in Fig. 10(b), generated spheroidite area fractions and mean sizes closely track real experimental trends with negligible deviation, demonstrating high-precision statistical replication.

Critically, the model generalizes beyond the training regime. Fig. 11 and Fig. B4 show that generated microstructures for unseen annealing temperatures largely accurately reproduce spheroidite shapes, distributions, size statistics, and area fractions. This good interpolative performance suggests our framework as a robust, data-efficient solution for predictive, parameter-aware microstructure generation, supporting accelerated materials design and discovery.

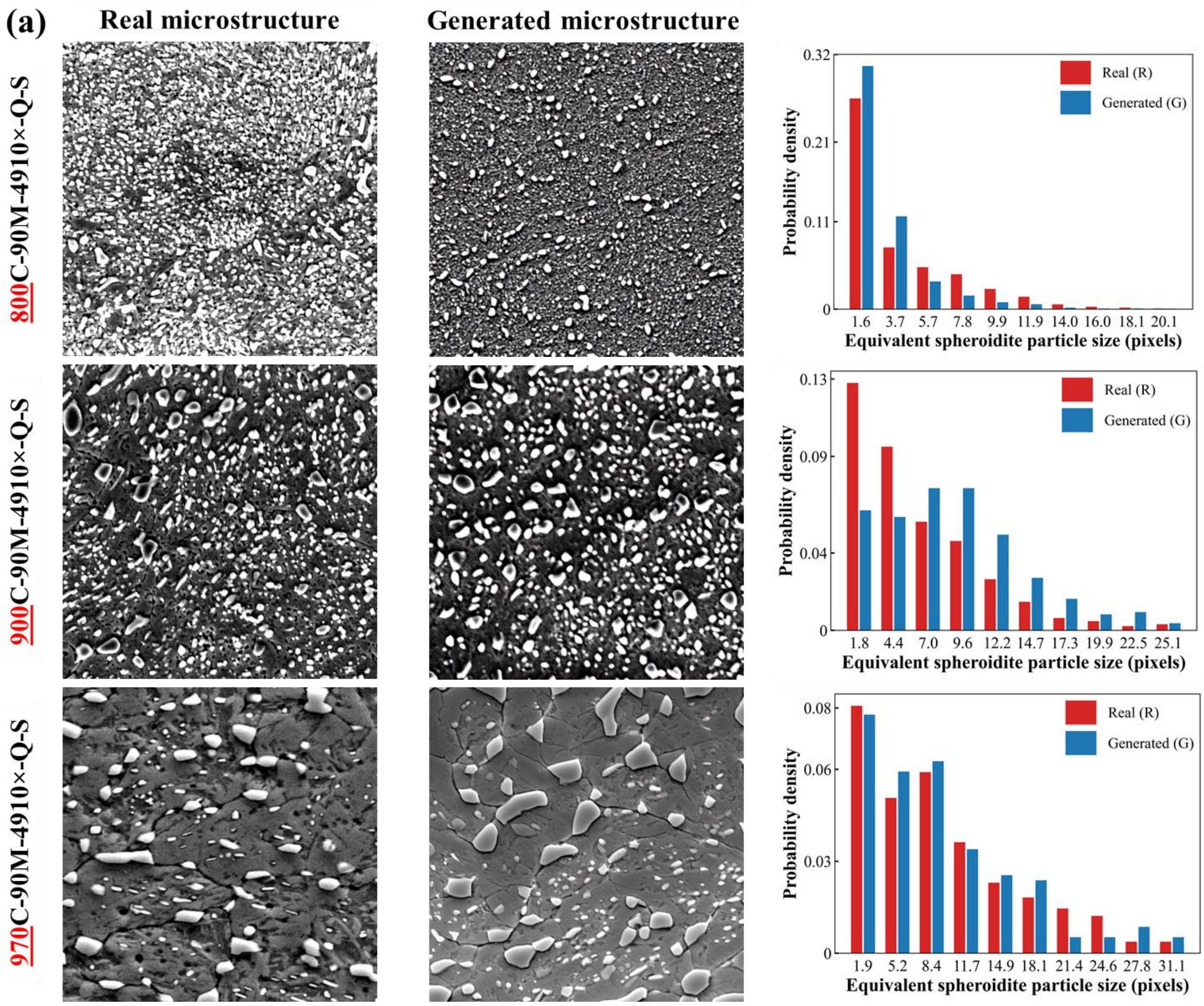

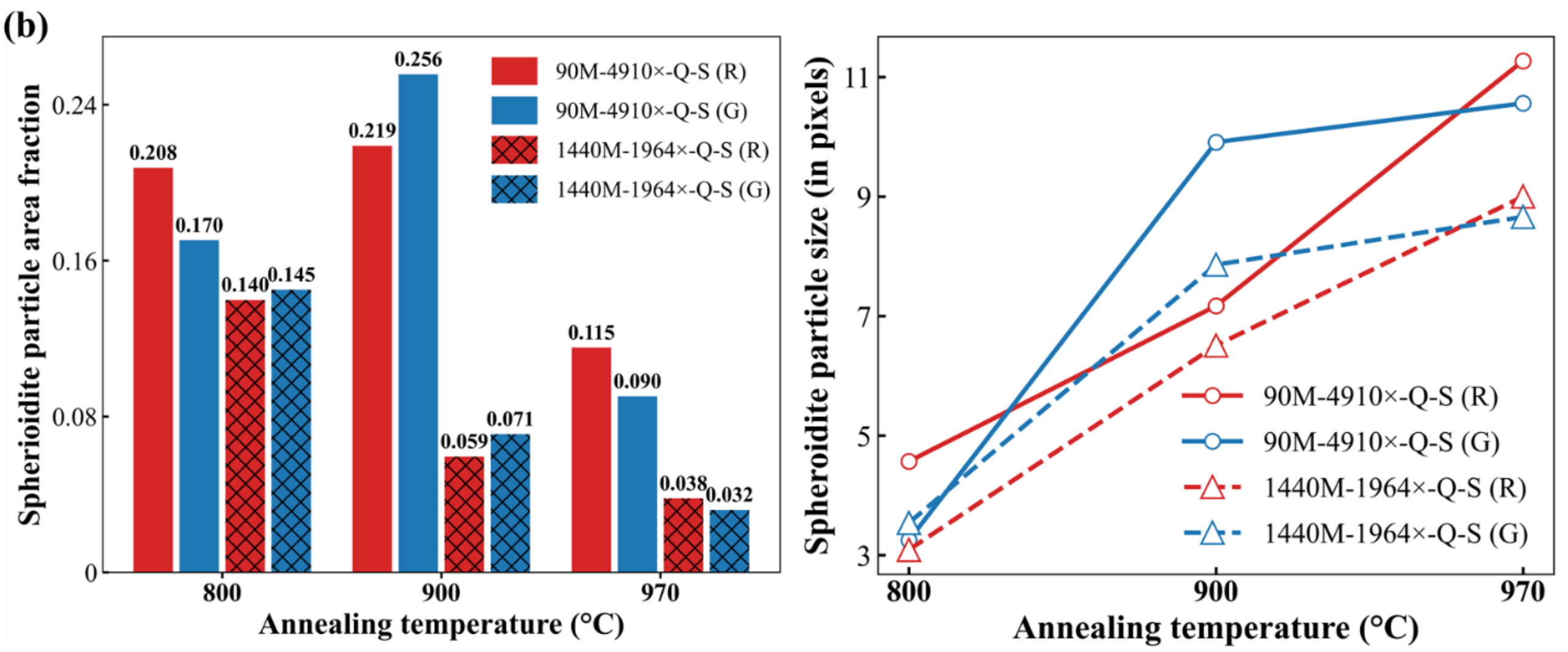

Fig. 10. Effect of temperature on microstructure generation under seen conditions: real and generated micrographs and corresponding spheroidite particle size distributions at three representative temperatures (800, 900, and 970 $^{o}$C) combined with (a) 90M-4910×-Q-S and (b) 1440M-1964×-Q-S, and (c) comparison of the spheroidite area fraction and mean particle size across these conditions.

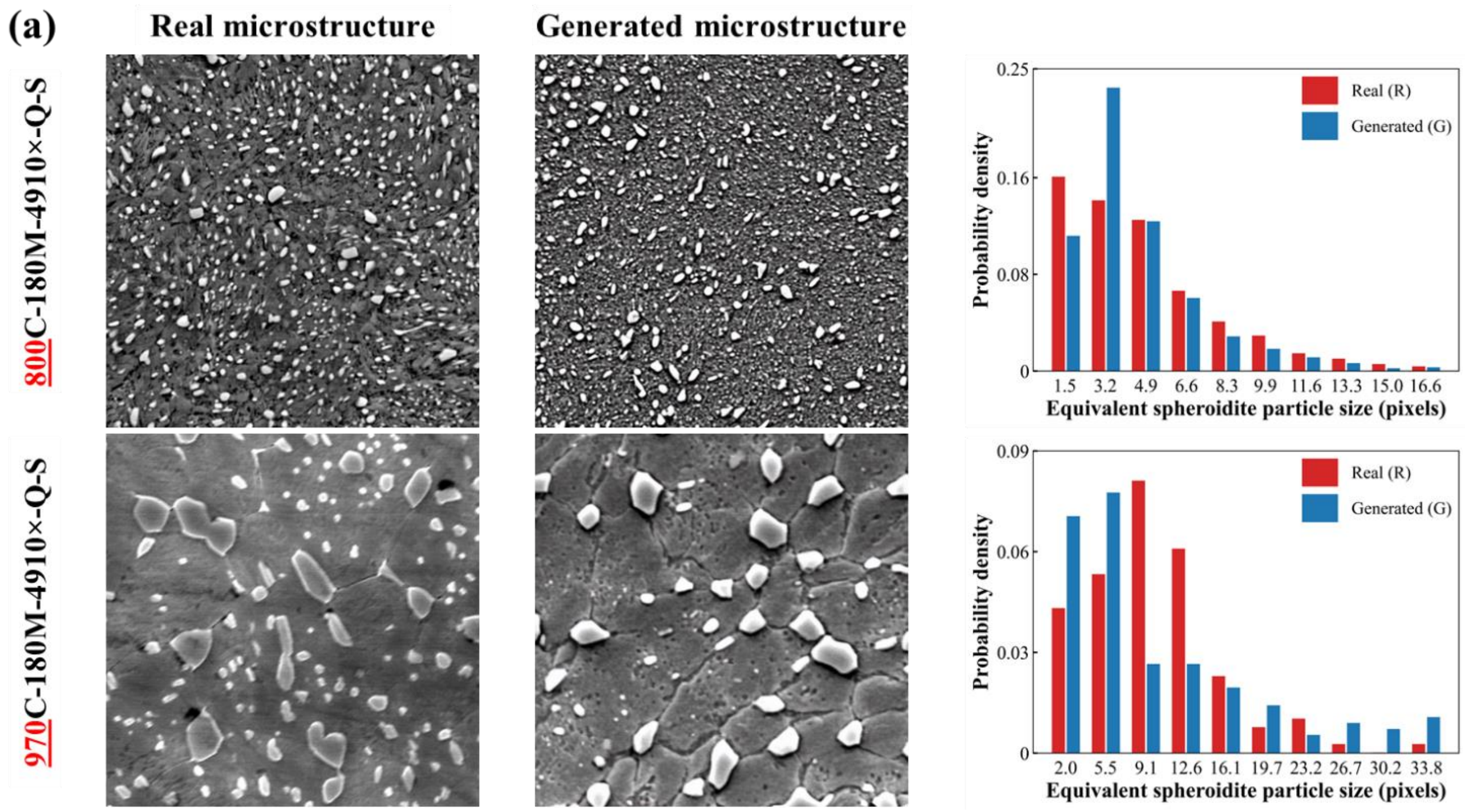

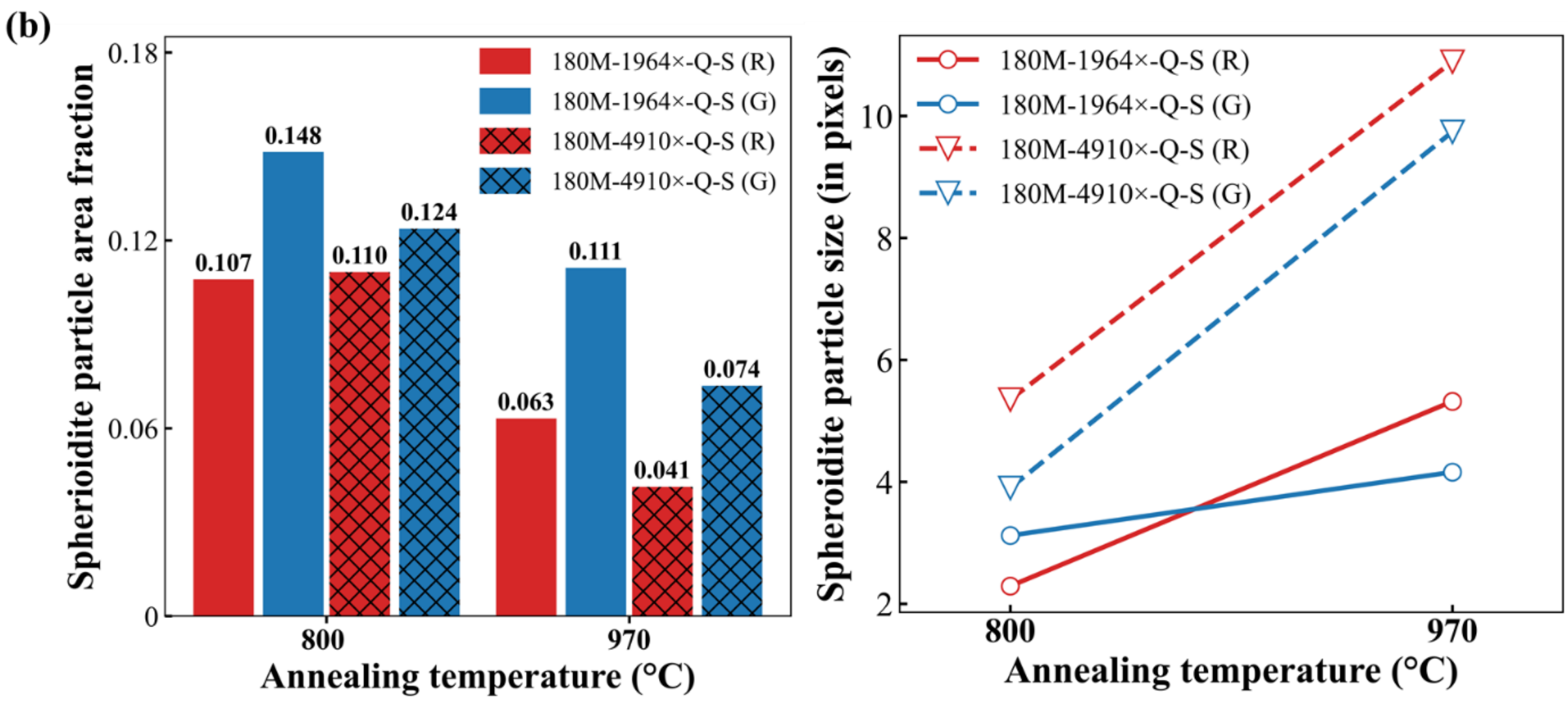

Fig. 11. Effect of temperature on microstructure generation under unseen conditions: real and generated micrographs and corresponding spheroidite particle size distributions at two representative temperatures (800 and 970 °C) combined with (a) 180M-4910×-Q-S and (b) 180M-1964×-Q-S, and (c) comparison of the spheroidite area fraction and mean particle size across these conditions.

*b. Examine various magnifications*

We further assess the model's capacity to generate microstructures across multiple imaging magnifications, a critical factor for capturing features at different length scales in the material. In the UHCS dataset, magnification serves as a conditional input, enabling the model to synthesize images at specific resolutions. Fig. 12(a) and Fig. B5 show real and generated spheroidite microconstituents at magnifications of 982×, 1964×, and 4910× under seen annealing conditions (800 °C for 5100 minutes and 800 °C for 1440 minutes, respectively). The model accurately reproduces microstructural features across scales, capturing the broad distribution of spheroidite particles at low magnification and resolving fine details at high magnification. Particle size distributions are in close agreement between real and generated images, though the model slightly underestimates the frequency of the largest particles at 4910×.

Quantitative analysis (Fig. 12(b)) reveals a direct correlation between magnification and the measured average spheroidite particle size. For example, samples annealed at 800 °C for 5100 minutes exhibit progressively larger sizes as magnification increases, consistent with the reduced field of view at higher magnifications. Notably, to our knowledge, no prior generative model has explicitly demonstrated preservation of this magnification-size scaling; our results thus go beyond purely visual validation.

For generalization, Fig. 13(a) and Fig. B6 evaluate an unseen annealing time (800 and 970 °C, 180 minutes) across magnifications. The model generates realistic microstructures with accurate particle shapes, spatial distributions, and size statistics across all scales. As confirmed by Fig. 13(b), the magnification-size correlation persists in both real and synthetic images, with minimal deviations.

These results demonstrate the parameter-aware Stable Diffusion model's robustness in simulating multi-scale microstructural characteristics, even under untrained conditions, paving the way for more quantitative comparisons of synthetic microstructures to real ones.

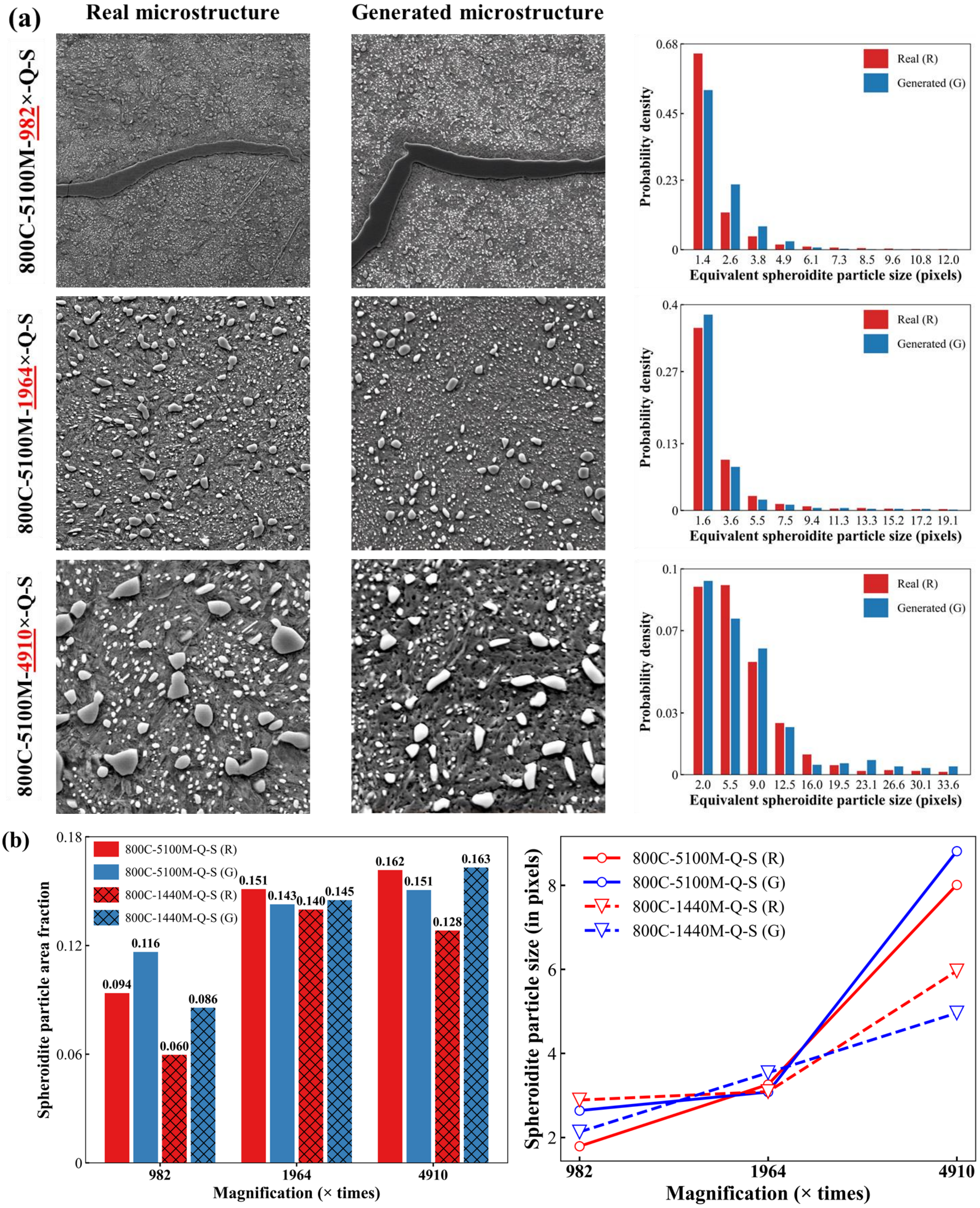

Fig. 12. Microstructure generation visualized at different imaging magnifications under seen conditions. Real and generated micrographs with corresponding spheroidite particle size distributions are shown at three representative magnifications (982×, 1964×, 4910×) combined with (a) 800C-5100M-Q-S, (b) 800C-1440M-Q-S; and (c) comparison of the spheroidite area fraction and average

particle size across conditions. Higher magnifications (smaller fields of view) show slightly larger measured particle sizes in both experimental and synthetic data, a trend successfully reproduced by the model.

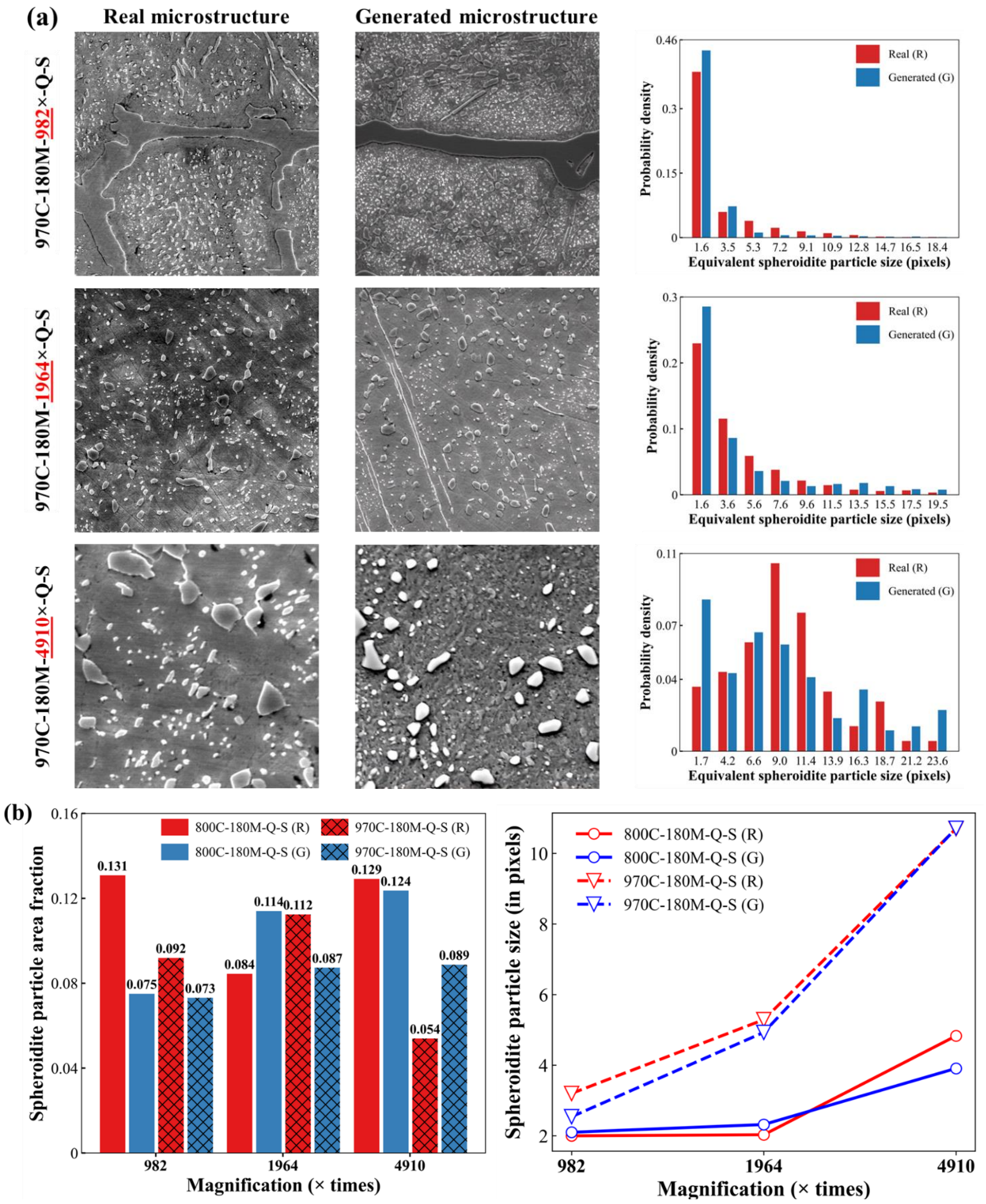

Fig. 13. Microstructure generation visualized at different imaging magnifications under unseen conditions. Real and generated micrographs with corresponding spheroidite particle size distributions are shown at three representative magnifications (982×, 1964×, 4910×) combined with (a) 970C-180M-Q-S, (b) 800C-180M-Q-S; and (c) comparison of the spheroidite area fraction and average particle size across conditions. The model closely replicates the real microstructures, as confirmed by both qualitative visualization and quantitative analysis.

*3.2.2.2. Spatial statistical functions*

*a. Visualization of spatial correlation functions*

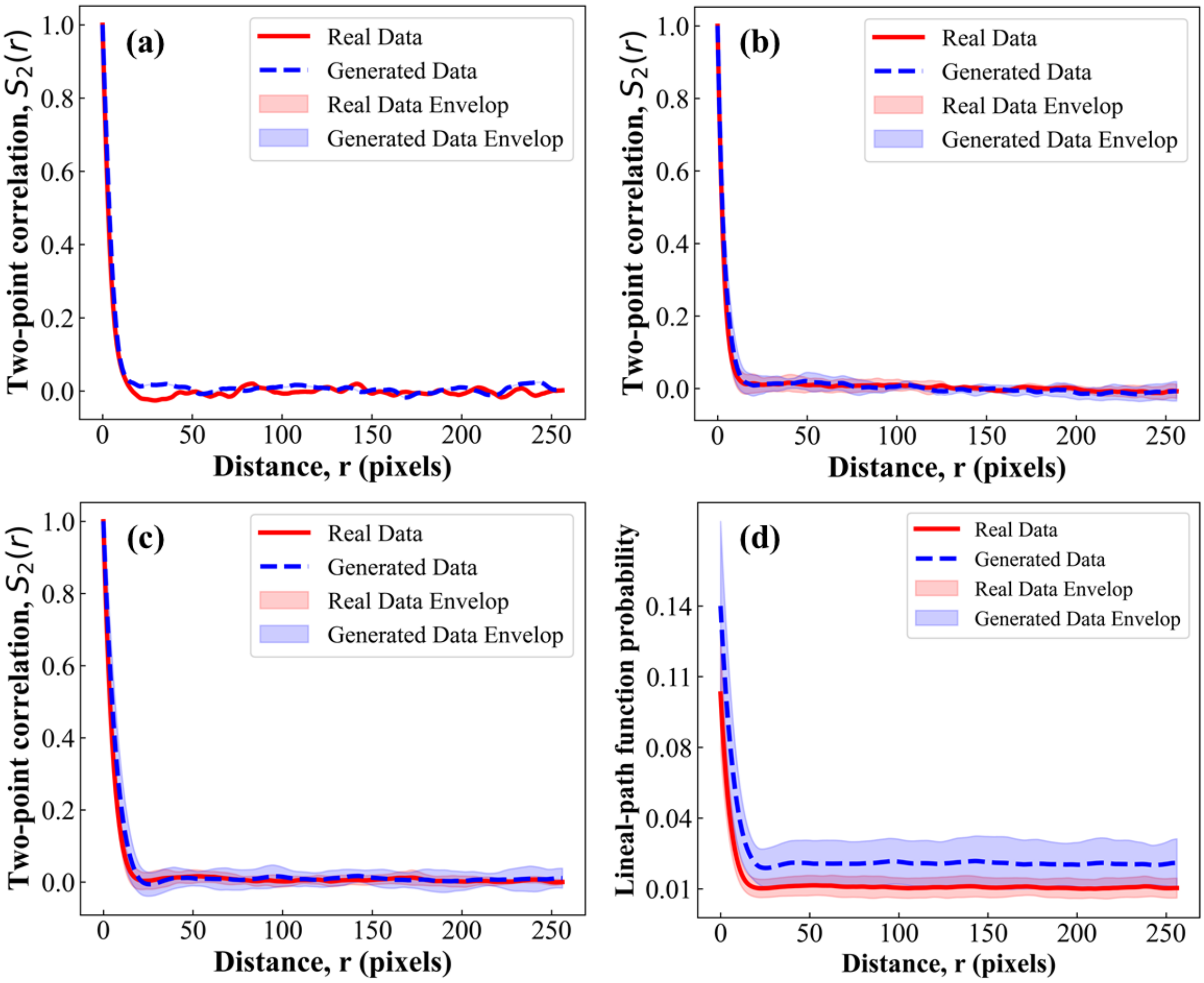

Fig. 14. Statistical comparison of correlation functions for real and generated micrographs, with uncertainty envelopes. (a-c) Two-point correlation for: (a) a seen (900C-90M-4910×-Q-S), (b) an unseen (800C-180M-4910×-Q-S), (c) an unseen (970C-180M-1964×-Q-S). (d) Lineal-path function for an unseen condition (970C-180M-1964×-Q-S).

Having confirmed close agreement in physical descriptors, we next assess statistical spatial functions to further evaluate the model's ability to capture the underlying spatial organization and phase connectivity. Fig. 14 illustrates a comparison between the correlation functions for real and generated microstructures. The mean two-point correlation curves of the generated images (blue, dashed lines) closely track those of the real data (red, solid lines) for both seen and unseen processing conditions, with only minor deviations. Particularly, at short distances ($r \approx 0$), the generated $S_2(r)$ matches the real curve almost exactly, confirming that the model accurately reproduces the phase fraction and nearest-neighbor arrangement of spheroidite. As $r$ increases, $S_2(r)$ for both real and generated data approaches zero, indicating that the synthetic images correctly capture the overall spatial randomness and phase proportions at large length scales. The close alignment of the mean correlation lines and the substantial overlap of their shaded uncertainty ranges demonstrate that key spatial statistics of the microstructure are preserved. Similarly, the lineal-path function results (Fig. 14(d)) show good agreement between experimental and generated samples. Across all distances, the probability of finding a continuous spheroidite path of length $r$ in the generated microstructures is nearly identical to that in the real ones, confirming that phase connectivity and continuity are faithfully reproduced. This high degree of agreement in both correlation functions indicates that the SD3.5-Large model produces visually realistic images while capturing the underlying spatial correlations essential for quantitative material characterization.

*b. Error analysis of spatial statistics*

We report normalized mean-squared errors ($\varepsilon$) for the two-point correlation $S_2(r)$ and lineal-path functions $L(r)$. The fine-tuned SD3.5-Large (full-text conditioning) attains the lowest errors of about 2.1% or less for $\varepsilon_{S_2}$ and 0.6% for $\varepsilon_L$ (Table 6). For context, prior microstructure reconstruction/generation studies have reported normalized descriptor-error about 3 to 5% as indicating close statistical agreement. We use this as a literature-reported reference range, not a universal standard; acceptable tolerances are application-dependent [15,47,52,53]. Our errors are well below this range and are consistent with the correlation-function envelopes (Fig. 14). Together with agreement in application-relevant observables (area fraction, size distributions) and preserved parameter-structure trends (temperature, annealing time, cooling-rate, magnification-size scaling), indicating fidelity sufficient for practical microstructure analysis and parameter-aware comparisons.

Table 6. Error metrics (%) for two-point correlation and lineal-path functions for baseline models and SD3.5-Large under different prompting conditions.

| **Condition** | StyleGAN3 | | DDPM | | **SD3.5 + DreamBooth + LoRA** | | | | | |
|---|---|---|---|---|---|---|---|---|---|---|
| | | | | | Parameter-only | | Text-only | | **Full-text** | |
| | $\varepsilon_{S_2}$ | $\varepsilon_L$ | $\varepsilon_{S_2}$ | $\varepsilon_L$ | $\varepsilon_{S_2}$ | $\varepsilon_L$ | $\varepsilon_{S_2}$ | $\varepsilon_L$ | $\varepsilon_{S_2}$ | $\varepsilon_L$ |
| **Seen processing** | | | | | | | | | | |
| 800C-1440M-1964×-Q-S | 7.40 | 3.94 | 5.56 | 2.83 | 5.34 | 1.21 | 1.68 | 0.28 | **1.40** | **0.28** |
| 900C-90M-4910×-Q-S | 9.97 | 9.69 | 3.71 | 2.89 | 2.78 | 1.62 | 3.22 | 1.46 | **1.93** | **0.21** |
| 970C-480M-1473×-Q-S | 9.77 | 4.18 | 2.38 | 1.83 | 2.35 | 1.34 | 4.26 | 0.43 | **1.70** | **0.16** |
| 800C-5100M-1964×-Q-S | 8.95 | 3.42 | 5.49 | 4.71 | 4.32 | 1.52 | 1.54 | 1.24 | **1.24** | 0.49 |
| **Unseen processing** | | | | | | | | | | |
| 800C-180M-4910×-Q-S | 9.73 | 4.71 | 8.48 | 3.13 | 3.62 | 0.56 | **1.10** | **0.41** | 1.14 | 0.59 |
| 800C-180M-1964×-Q-S | 6.49 | 5.14 | 3.45 | 1.02 | 4.31 | 0.86 | 1.68 | **0.36** | **1.38** | 0.57 |
| 970C-180M-982×-Q-S | 3.34 | 5.34 | 2.19 | 4.11 | 1.59 | **0.20** | 1.42 | 0.23 | **1.22** | 0.30 |
| 970C-180M-1964×-Q-S | 12.09 | 4.77 | 5.80 | 3.39 | 2.39 | 0.55 | 2.91 | 1.39 | **2.09** | **0.13** |

Table 6 compares StyleGAN3, DDPM, and SD3.5-Large under parameter-only, text-only, and full-text prompts. Full-text-prompts yield the lowest errors across all conditions, and even parameter-only prompt outperforms the baselines. Across the four seen settings, SD3.5-Large improves on both StyleGAN3 and DDPM for $\varepsilon_{S_2}$ and $\varepsilon_L$. In addition, across the four unseen selections, it exceeds DDPM in three of four for $\varepsilon_{S_2}$ and all four for $\varepsilon_L$. For example, at 970C-480M-1473×-Q-S, full-text model achieves $\varepsilon_{S_2}$=1.70% and $\varepsilon_L$=0.16% compared with parameter-only 2.35% and 1.34%, DDPM 2.38% and 1.83%, and StyleGAN3 9.77% and 4.18%. These comparisons reveal that SD3.5-Large performs better than GAN and DDPM baselines for this task and that combining textual context with continuous and categorical variables is essential for statistical fidelity and generalization.

### 3.3. External validation on Aachen-Heerlen

We evaluate cross-dataset robustness on the Aachen-Heerlen (A-H) dataset [35], whose condition schema (continuous and categorical) differs from UHCS. We reuse the SD3.5-Large architecture and training recipe without modification and keep all UHCS hyperparameters; performance is summarized in Table 7(a). For context relative to the in-domain UHCS baseline (Table 2, full-text), Table 7(a) illustrates higher CMMD on A-H, with CLIP score/SSIM/LPIPS of similar magnitude, Precision/Recall are lower. These differences likely reflect A-H's small, more fragmented martensite-austenite (MA) regions, combined with the fact that we kept the UHCS hyperparameters unchanged. For comparability with the original A-H benchmark, our segmentation pipeline achieves an IoU of 0.487 for the MA islands, which is 39.14% relative improvement over original 0.35 [35]. Qualitative examples under unseen conditions (Fig. 15) show close visual agreement between generated and real micrographs; labels (e.g., 400C-9mm-V-P-2) encode temperature and distance (continuous), and vertical (V), angle (P), and chemical composition type (categorical). Distributional checks corroborate these trends: per-type MAE in MA area fraction is low (0.5-5.0 percentage points) and area fraction of MA (Table 7(a), Fig. D1), generally within one standard deviation of the real-data ordering a modest upward bias for a few types. The t-SNE embedding of features for real and generated micrographs shows substantial manifold overlap, indicating strong distributional alignment on this dataset.

Table 7. Performance on Aachen-Heerlen (A-H) dataset. (a) Model performance of image-quality and segmentation metrics. (b) Per-type mean absolute error (MAE) in MA area fraction reported in percentage points (pp). MA-IoU is reported as ours/original A-H benchmark.

(a) Model performance metrics

| **Model** | **CMMD** (↓) | **CLIP Score** (↑) | **SSIM** (↑) | **LPIPS** (↓) | **Precision** (↑) | **Recall** (↑) | **MA-IoU** (*ours / orig.*) |
|---|---|---|---|---|---|---|---|
| SD3.5-Large (full-text) | 1.185 | 0.302 | 0.778 | 0.415 | 0.578 | 0.647 | 0.487 / 0.35 |

(b) Per-type MAE of MA area fraction

| **Type** | **1** | **CFB2** | **5** | **10** | **11** | **2** | **3** | **8** | **12** |
|---|---|---|---|---|---|---|---|---|---|
| MAE (pp) ↓ | 2.53±3.77 | 1.79±1.28 | 4.45±5.20 | 4.96±5.88 | 3.71±5.38 | 2.06±2.05 | 0.54±0.46 | 0.57±0.55 | 3.18±5.38 |

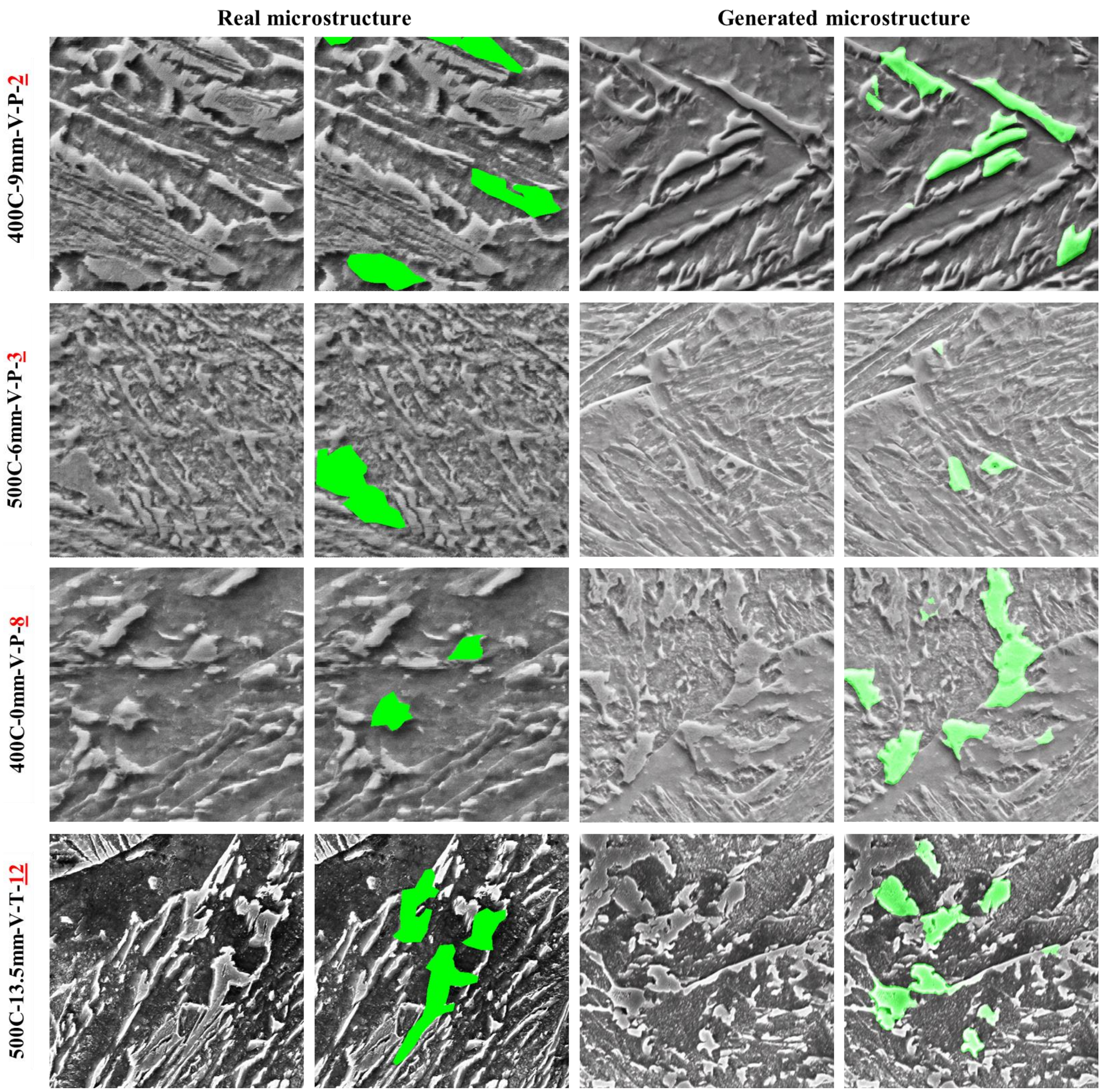

Fig. 15. External validation on Aachen-Heerlen with data-specific labels under unseen conditioning.

## 4. Conclusions

This study demonstrates that large text-to-image diffusion models can be adapted to materials science by directly encoding processing parameters as model conditioning inputs. We inject multiple continuous and categorical parameters into the CLIP token stream of Stable Diffusion 3.5-Large for conditioning. In addition, we fine-tune only the model's adapter layers via DreamBooth and LoRA. This strategy enables controllable microstructure generation while requiring updates to only a small fraction of the model's weights.

On ultra-high-carbon steel, the method consistently outperforms StyleGAN3, DDPM, and SDXL on image-quality and structural metrics. A VGG-16 U-Net (85.7% mIoU, 97.1% accuracy) provides segmentation masks for computing physical descriptors and spatial statistics. Generated microstructures match real samples in particle-size distributions and area fractions, and achieve normalized two-point and lineal-path errors below 2.1% and 0.6%, respectively, across both seen and unseen processing conditions. Notably, the approach generalizes to the Aachen-Heerlen dataset simply by swapping out the parameter schema. No architectural changes to the model are required for this transfer.

Overall, our results indicate that parameter-aware diffusion provides a scalable, data-efficient route to explore process-structure relationships and to generate synthetic micrographs for downstream tasks (classification, detection, segmentation). While full-text prompts yield the highest fidelity, we find that parameter-only prompts already produce useful images for design exploration. We expect the fidelity gap between these two prompting modes to narrow as larger, more balanced datasets become available and as parameter-encoding techniques are strengthened.

Our work is limited to two steel systems and two-dimensional micrographs. Future efforts will extend to other materials families and imaging modalities. We also plan to incorporate physics-guided priors and property-conditioning, address class imbalance with expanded training data and attention mechanisms, and refine segmentation with expert-validated labels. These improvements are aimed at broadening applicability and reducing reliance on costly experimentation.

## Acknowledgements

This research was supported by Technology Innovation Program (P0022331) through the Korea Institute for Advancement of Technology funded by the Korea Ministry of Trade, Industry and Energy and by Nano and Materials R&D program (RS-2024-00451579) through the Korea Science and Engineering Foundation funded by the Ministry of Education, Science and Technology.

## Ethics approval

Not Applicable.

## Consent to participate

Not Applicable.

## Consent for publication

Not Applicable.

## CRediT authorship contribution statement

**Hoang Cuong Phan:** Conceptualization, Investigation, Methodology, Software, Data curation, Writing - original draft. **Minh Tien Tran:** Investigation, Methodology. **Chihun Lee:** Investigation, Methodology. **Hoheok Kim:** Investigation, Methodology. **Sehyeok Oh:** Investigation, Methodology. **Dong-Kyu Kim:** Investigation, Methodology. **Ho Won Lee:** Conceptualization, Methodology, Supervision, Funding acquisition, Review & Editing.

## Declaration of competing interest

The authors declare that they have no known competing financial interests or personal relationships that could have appeared to influence the work reported in this paper.

## Code availability

Code available on request from the authors

## Declaration of generative AI and AI-assisted technologies in the writing process

During the preparation of this work the authors used ChatGPT in order to improve readability. After using this tool/service, the authors reviewed and edited the content as needed and take full responsibility for the content of the publication.

## Appendix A. Extended methodology

### A1. Dataset preparation and preprocessing

Table A1. UHCS dataset splits and process parameter ranges

| Split | No. images | Annealing temperature (°C) | Annealing time (minutes) | Cooling methods | Magnification (×) |
|---|---|---|---|---|---|
| **UHCS [28], 598 images used for microstructure generation** | | | | | |
| Training | 521 | 700,750, 800, 900, 970, 1000, 1100 | 5, 60, 90, 180, 480, 1440, 2800, 5100 | 650-1H, AR, Q, FC | 49, 64, 74, 98, 118, 147, 246, 344, 491, 638, 786, 982, 1178, 1473, 1964, 2455, 3437, 4910, 7856, 9820, 11785, 19641, 34372 |
| Validation (Seen) | 23 | 800, 970 | 90, 480 | Q | 982, 1473, 1964, 3437, 4910 |
| Validation (Unseen) | 54 | 800, 970 | **180** | Q | 982, 1473, 1964, 3437, 4910 |
| **Subset of UHCS [31], 24 labeled images used for semantic segmentation** | | | | | |

✓ *650-1H: held at 650 °C for 1 hour, FC: furnace cooled, Q: quenched, AR: air cooled*
✓ *Seen conditions: combinations of process parameters + magnification already present in the training set.*
✓ *Unseen conditions: combinations of process parameters + magnification not present in the training set (in this paper, annealing time of 180 minutes are set aside for evaluating generalization capability).*

The annealing time of 180 minutes was exclusively reserved for the unseen validation set for two main reasons. First, given the limited size of the UHCS dataset, selecting this specific time point enabled the creation of diverse validation samples by combining it with multiple temperatures, cooling methods, and magnifications to ensure enough data coverage without overlapping the training set. Second, this choice was intentionally aligned with the UHCS subset used for semantic segmentation. This compatibility allows direct comparison between real and generated microstructures using quantitative metrics.

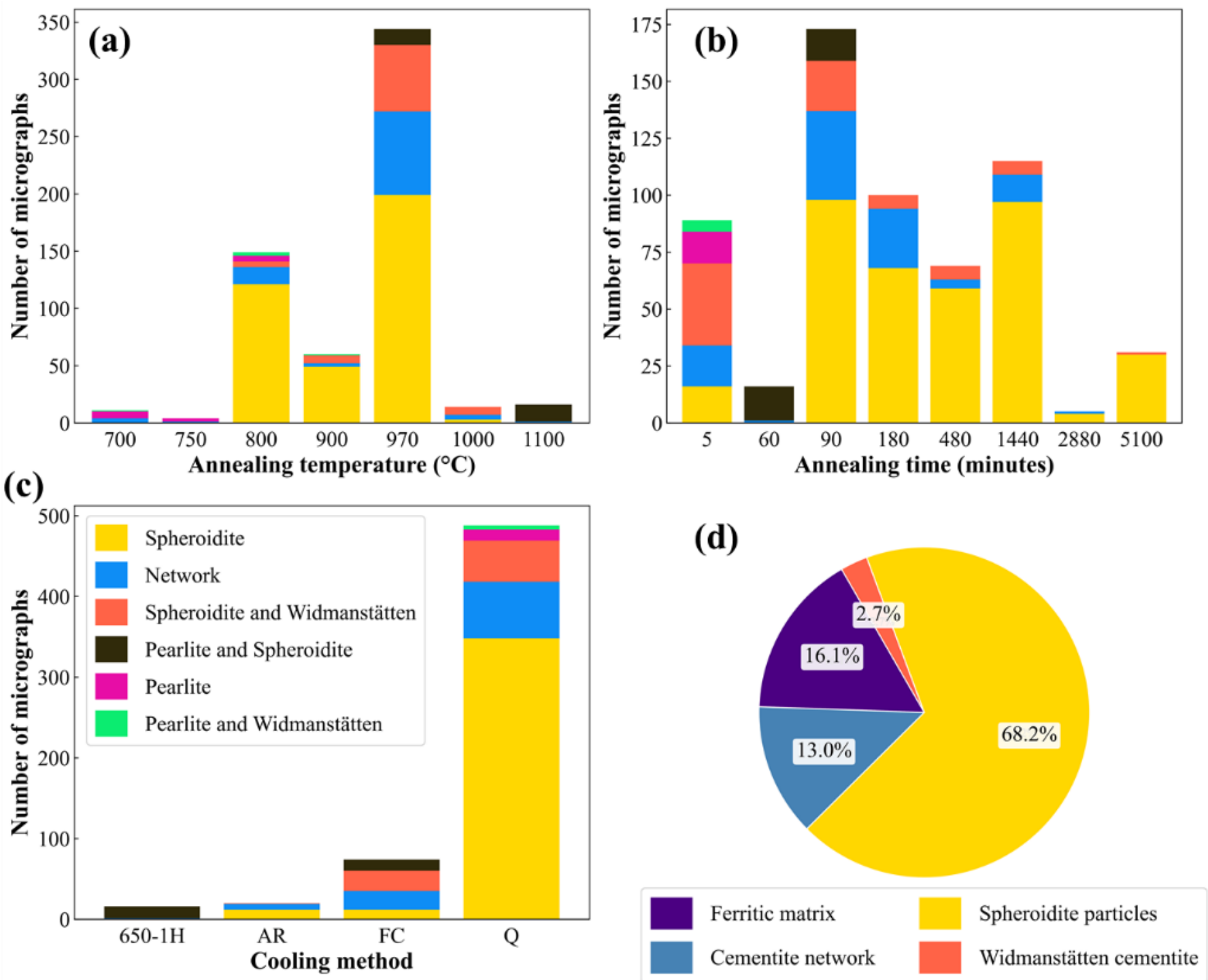

Fig. A1. (a-c) Number of micrographs by annealing temperature, annealing time, and cooling method in the UHCS dataset. (d) Class distribution labeled UHCS subset.

### A2. Evaluation metrics

To assess the quality and fidelity of generated microstructures, we use a set of metrics tailored to text-guided image generation and segmentation. For image generation, we compute the CLIP score with CLIP ViT-L/14 [54], i.e., the cosine similarity between image and text embeddings [55,56]. Higher CLIP scores indicate stronger alignment between generated images and their process-conditioned prompts and correlate with human judgment.

While Fréchet Inception Distance (FID) [23] is widely used, its reliance on very large sample sizes (commonly 50000, some cases requiring 10000 images) and sensitivity to biased embeddings [57] make it unsuitable for the data-scarce UHCS setting. We therefore adopt CLIP-Maximum Mean Discrepancy (CMMD) [57], which provides reliable estimates with small datasets and has outperformed FID in materials-science applications. CLIP score and CMMD serve as our primary metrics for fidelity and distributional alignment.

As secondary metrics, we report Precision and Recall [58] to assess sample fidelity and coverage, respectively. We also include Structural Similarity Index (SSIM) [59] and Learned Perceptual Image Patch Similarity (LPIPS) [60] to capture pixel-level structural and perceptual differences, providing a holistic assessment of visual quality.

### A3. Experiment setup

For microstructure generation, we fine-tune SDXL and SD3.5 (Medium and Large) models using DreamBooth and LoRA on the UHCS dataset at a resolution of 512 × 512. To enhance generalization during training, we apply data augmentation techniques, including random horizontal flips and random cropping. The AdamW optimizer [24,61] is employed with a constant learning rate, without warmup or additional regularization, following [61]. Mixed-precision training with bfloat16 (bf16) is used to optimize computational efficiency on GPU hardware. All remaining hyperparameters are set to their default values as specified in [24,29], with a full list provided in Table B1. All training is conducted on a single NVIDIA A100-80 GB GPU, except for SDXL are trained on four NVIDIA RTX A6000 GPUs. Inference uses a single NVIDIA RTX A6000 GPU with the FlowMatchEulerDiscreteScheduler sampler. All experiments were performed with PyTorch 2.1.0, CUDA 11.8, and Diffusers 0.33.0. Random seeds were fixed to ensure reproducibility.

For microstructure segmentation, to mitigate overfitting due to the small dataset, we apply various regularization techniques, including group normalization (8 groups), dropout (rate 0.3), weight decay (0.0005). Data augmentation includes random horizontal flips (probability 0.8), random cropping, rotations (0-45$^{o}$, probability 0.8), and random adjustments to brightness and contrast (range 0.2 to 0.8). The AdamW optimizer is used, with separate learning rates for the encoder and decoder. Segmentation experiments are executed on a single NVIDIA RTX A6000 GPU.

## Appendix B. Extended generation results

### B1. Hyperparameter tuning

Table B1. Hyperparameter configurations for various fine-tuning strategies on SDXL and SD3.5 (Medium and Large) models. We list parameters for: (i) LoRA-only, (ii) DreamBooth + DoRA, (iii) DreamBooth + LoRA, all applied to the UHCS dataset.

| **Hyperparameters** | **Selected values of** | | | | |
|---|---|---|---|---|---|
| | **SDXL** with | | | **SD3.5** with | |
| | LoRA *(1)* | DreamBooth and DoRA *(2)* | DreamBooth and LoRA *(2)* | DreamBooth and LoRA (Medium) | DreamBooth and LoRA (Large) *(4)* |
| Rank | 8 | 8 | 4 | 128 | 256 |
| Learning rate | $1 \times 10^{-5}$ | $1 \times 10^{-4}$ | $8 \times 10^{-5}$ | $5 \times 10^{-5}$ | $8 \times 10^{-5}$ |
| Text encoder learning rate | - | $5 \times 10^{-6}$ | $5 \times 10^{-6}$ | $5 \times 10^{-6}$ | $8 \times 10^{-6}$ |
| Batch size *(3)* | 10 | 2 | 8 | 12 | 6 |
| Sampling steps | 250 | 50 | 250 | 50 | 50 |
| Classifier-free guidance | 6 | 6 | 4 | 3 | 2 |

*(1) Update U-Net + w/o update text encoders; (2) Update U-Net + with update text encoders; (3) batch size per one GPU (NVIDIA GTX A6000); (4) The SD3.5-Large model was trained on one GPU (NVIDIA A100-80 GB)*

Table B2. Hyperparameters for StyleGAN3 and DDPM.

| Hyperparameter | StyleGAN3 | DDPM |
|---|---|---|
| Learning rate | Generator: 0.0015, discriminator: 0.0001 | 0.001 |
| Batch size | 12 | 16 |
| Mapping layers | 8 | - |
| R1 regularization gamma | 5 | - |
| Noise schedule | - | linear |
| Timesteps | - | 1000 |
| Prediction type | - | epsilon |
| Optimizer | Adam | AdamW |
| Sampling steps | - | 50 (DDIM) |
| Classifier-free guidance (CFG) | - | 1.5 |

Table B3. Breakdown of total and trainable parameters in SD3.5-Large components (in millions).

| **Component** | **Total parameters (M)** | **Trainable parameters (M)** | **Trainable (%)** |
|---|---|---|---|
| Multimodal diffusion transformer (MM-DiT): 8056.6 M; LoRA adapters: 377.3M | 8 433.9 | 377.3 | 4.5 |
| CLIP-L/14 | 124 | 18.9 | 13.2 |
| CLIP-G/14 | 695 | 83.9 | 10.8 |
| T5-XXL | 4 762 | 0 | 0 |
| VAE | 83.8 | 0 | 0 |
| Parameter encoder (categorical + continuous) | 26.9 | 26.9 | 100 |
| **Overall summary** | 14 228 | 507 | **3.6** |

## B2. Additional generation experiments

B2.1. Extended qualitative comparison: Full-text versus text-only

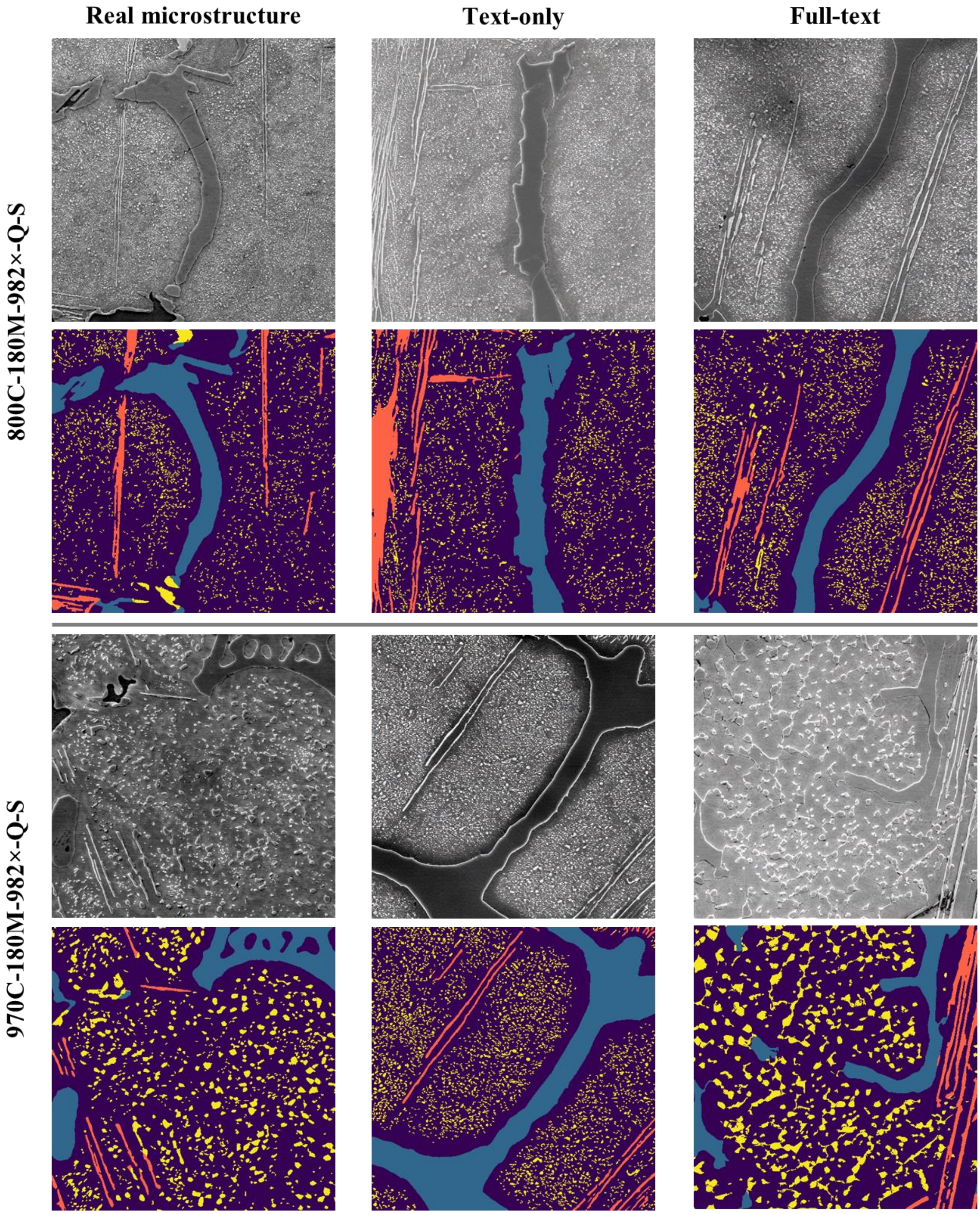

Fig. B1. Two examples illustrating the impact of parameter-aware conditioning on generation fidelity compared to models without this feature.

B2.2. Ablation study: benchmarking other Stable Diffusion models

Using the full-text conditioning scheme, we compared stable diffusion variants. Our fine-tuned

SD3.5-Large demonstrates superior fidelity and diversity in microstructure generation, outperforming all other configurations across standard image quality metrics. Specifically, it achieved the lowest LPIPS (0.429), highest Precision (0.921) and Recall (0.789), and the highest CLIP-Score (0.321), as summarized in Table B4. While SDXL fine-tuned with DreamBooth + LoRA attained the lowest CMMD, the SD3.5-Large model consistently delivered the best results for perceptual similarity, feature coverage, and overall alignment with real microstructure distributions. These outcomes indicate that SD3.5-Large not only generates images perceptually closest to real samples, but also maintains broad coverage of the underlying data distribution.

Table B4. Quantitative comparison of SDXL and SD3.5 variants. SDXL was fine-tuned using only LoRA, DreamBooth + DoRA, and DreamBooth + LoRA, while SD3.5 has Medium and Large variants, both fine-tuned with DreamBooth + LoRA.

| **Model configuration** | **CMMD** (↓) | **CLIP Score** (↑) | **SSIM** (↑) | **LPIPS** (↓) | **Precision** (↑) | **Recall** (↑) |
|---|---|---|---|---|---|---|
| SDXL + LoRA | 0.419 | 0.312 | **0.668** | 0.470 | 0.882 | 0.658 |
| SDXL + DreamBooth + DoRA | 0.467 | 0.316 | 0.613 | 0.473 | 0.868 | 0.605 |
| SDXL + DreamBooth + LoRA | **0.357** | 0.317 | 0.651 | 0.449 | 0.895 | 0.610 |
| SD3.5-Medium + DreamBooth + LoRA | 0.572 | 0.320 | 0.611 | 0.432 | 0.816 | 0.684 |
| **SD3.5-Large + DreamBooth + LoRA** | 0.538 | **0.321** | 0.654 | **0.429** | **0.921** | **0.789** |

We also examined the fidelity of generated microstructures from the top-performing SDXL and SD3.5 models fine-tuned using DreamBooth and LoRA. Microstructure images generated by SDXL exhibited several deficiencies in reproducing fine microstructural details (Fig. B2), likely due to reliance exclusively on CLIP text encoders for conditioning. For example, Widmanstätten cementite laths were often missing, appeared irregular, overly thin, or fragmented. Additionally, SDXL frequently failed to clearly depict denuded zones (carbide-depleted regions) [28,62], causing ambiguity between ferritic matrix and cementite networks. Furthermore, SDXL struggled to generate very fine spheroidite particles, rendering low-density spheroidite regions indistinct from the matrix phase. These issues made it difficult to differentiate specific microconstituents in SDXL's images. In contrast, SD3.5-Large faithfully reproduced all these critical microstructural features.

Based on the above quantitative and qualitative comparisons (Fig. B2 and Table B4), we selected SD3.5-Large for all subsequent microstructure generation experiments. We attribute SD3.5-Large's superior fidelity to (*i*) richer textual conditioning via dual CLIP encoders together with a large T5-XXL text model, and (*ii*) the higher-capacity MM-DiT image decoder, which together outperform the two CLIP text encoders and U-Net pipeline of SDXL (see Section 2.3 for architectural details). This combination yielded more structurally accurate and detailed synthetic microstructures, particularly for features requiring fine-grained control of phase morphology and spatial arrangement.

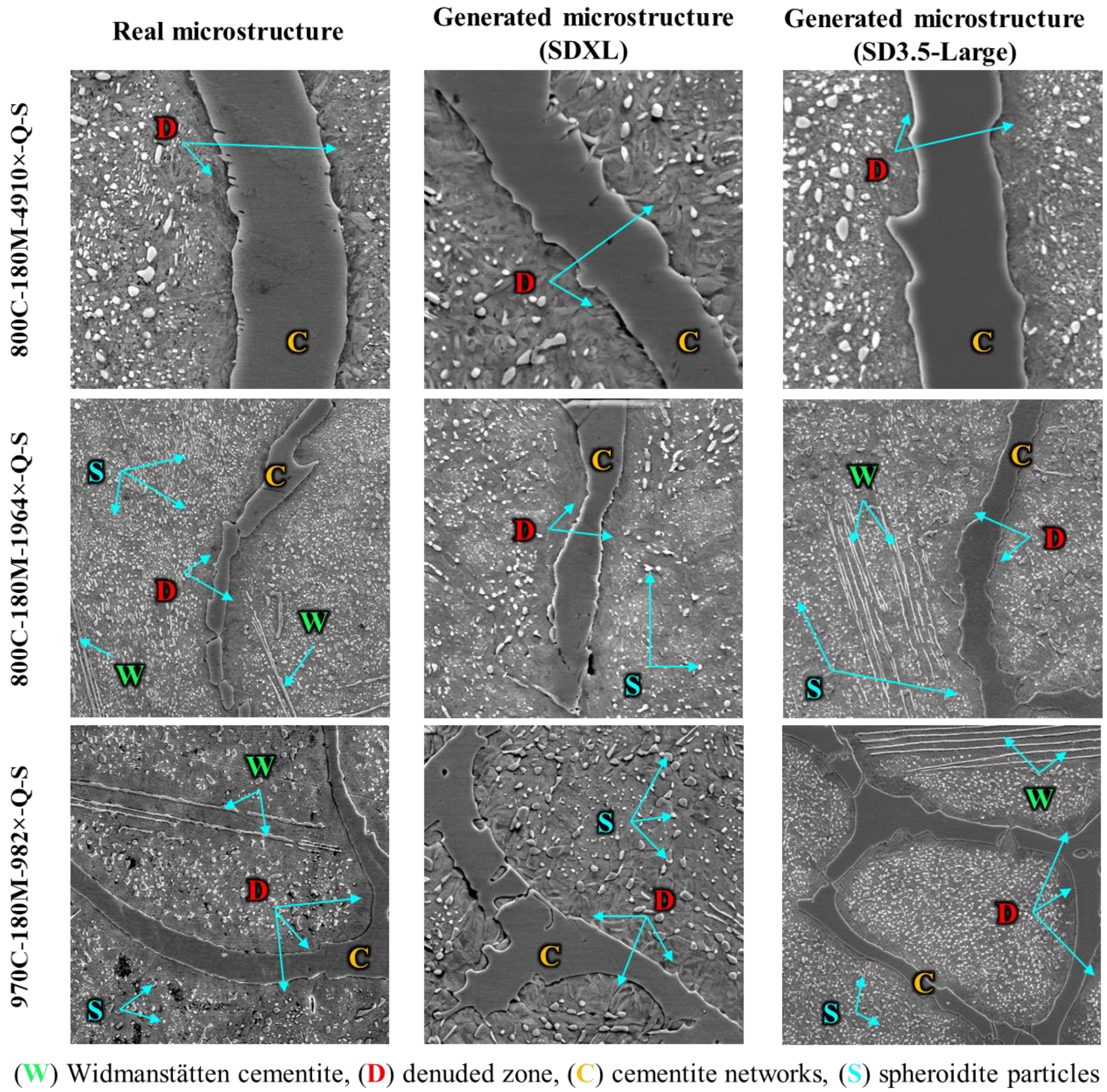

Fig. B2. Comparative analysis of microstructure generation: Deficiencies in SDXL (middle) versus enhanced fidelity in SD3.5-Large (right) across three representative examples. SDXL outputs exhibit critical inaccuracies: (**W**) Widmanstätten cementite laths are missing or fragmented; (**D**) denuded zones (carbide-depleted regions) are not clearly rendered; (**C**) proeutectoid cementite networks are incomplete; and (**S**) fine spheroidite particles are poorly represented. In contrast, SD3.5-Large accurately reproduces all microconstituents, demonstrating superior generative capability.

B2.3. Effects of individual parameters on generation

*a. Effect of annealing temperature*

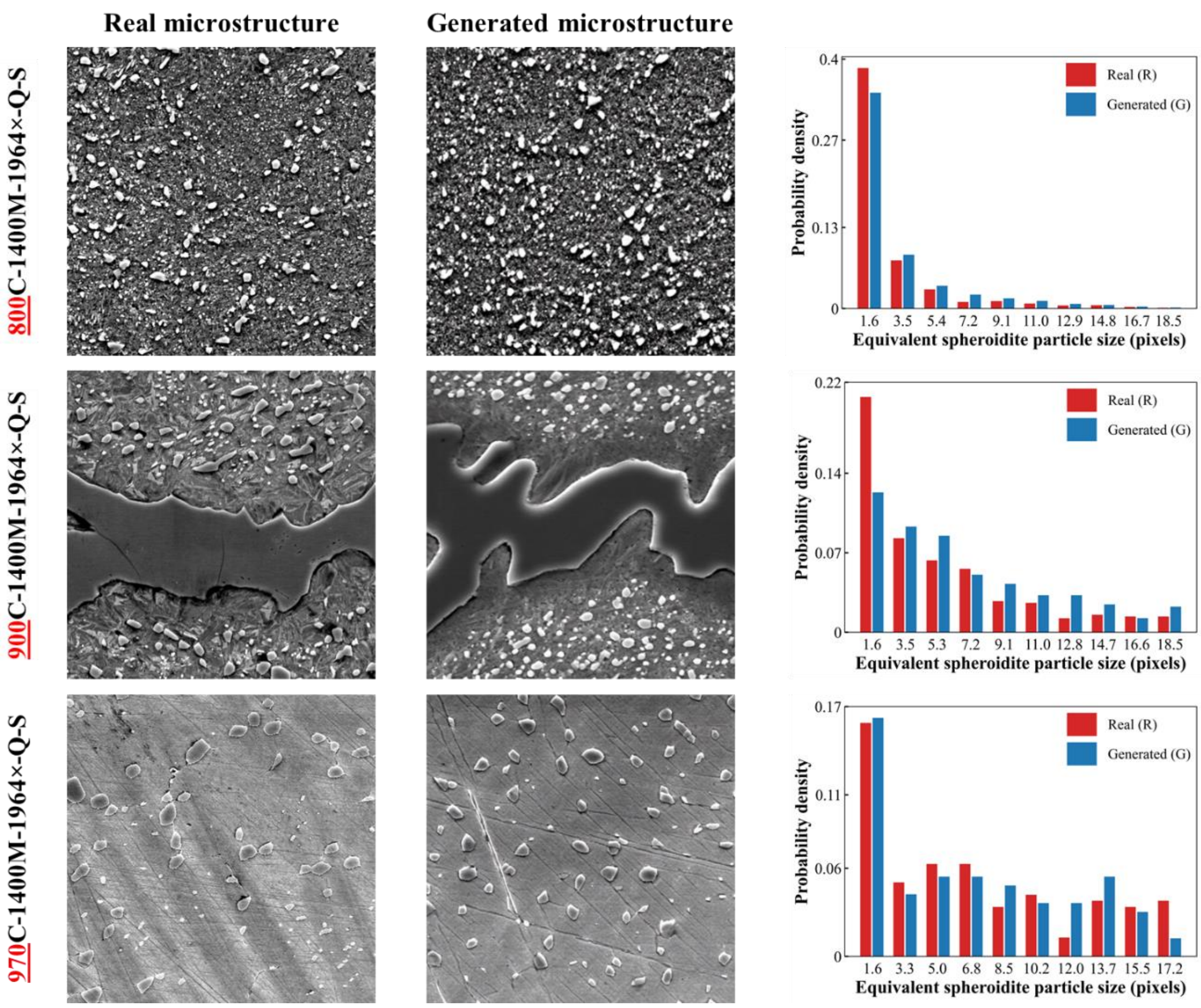

Fig. B3. Effect of temperature on microstructure generation under seen conditions: real and generated micrographs and corresponding spheroidite particle size distributions at three representative temperatures (800, 900, and 970 ºC) combined with 1440M-1964×-Q-S.

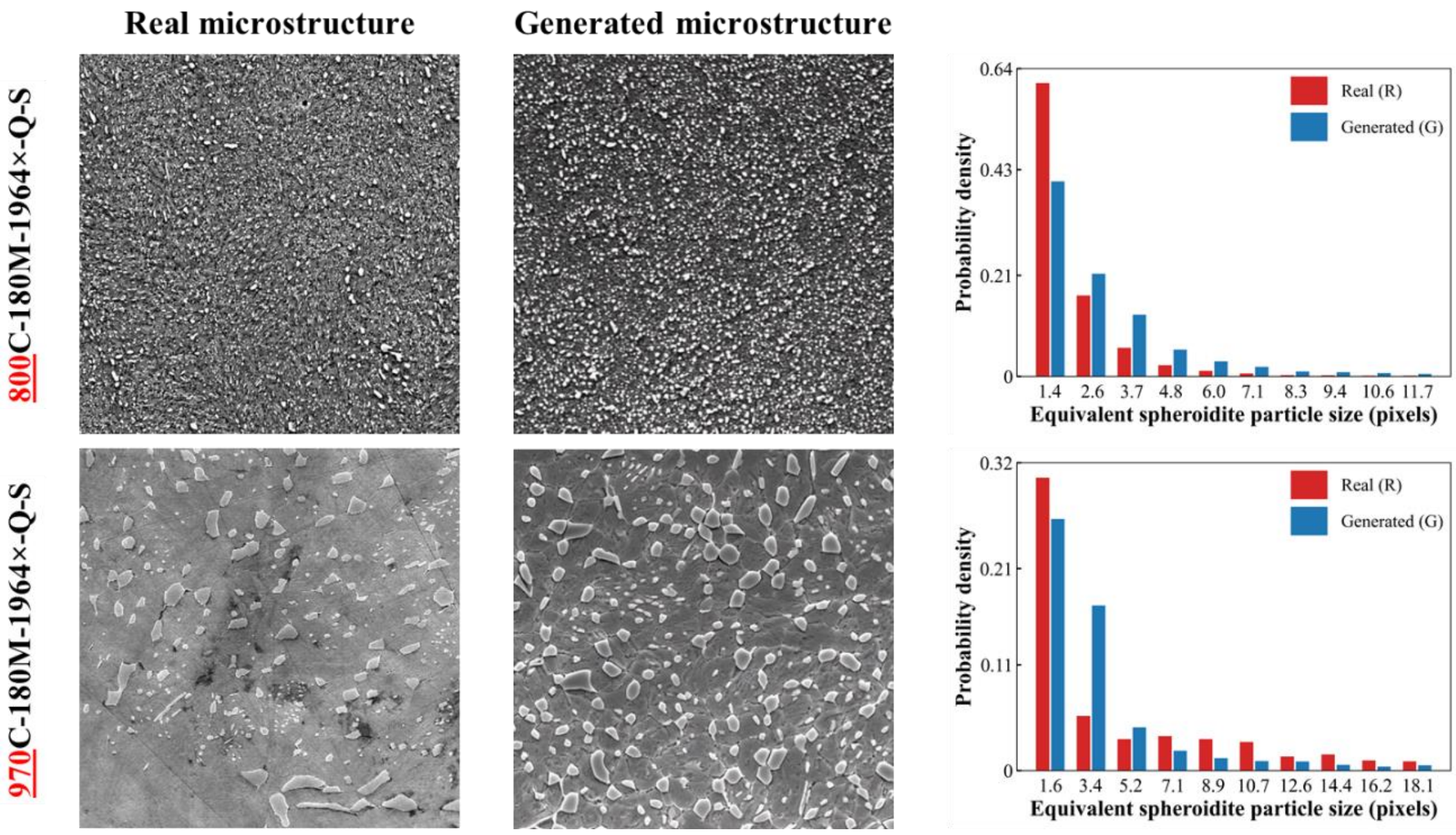

Fig. B4. Effect of temperature on microstructure generation under unseen conditions: real and generated micrographs and corresponding spheroidite particle size distributions at two representative temperatures (800 and 970 ºC) combined with 180M-1964×-Q-S.

*b. Magnification scaling*

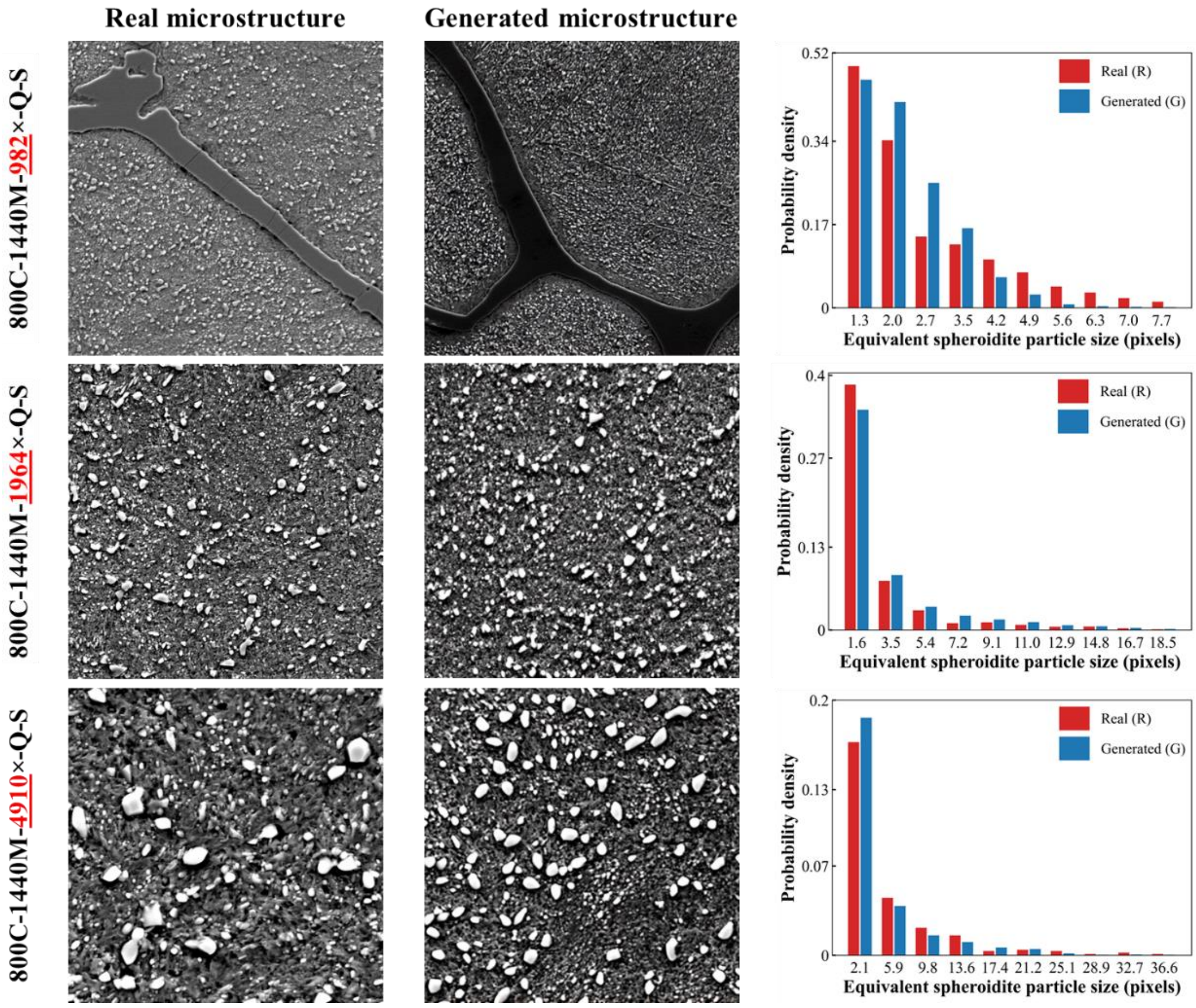

Fig. B5. Microstructure generation visualized at different magnifications under seen conditions. Real and generated micrographs with corresponding spheroidite particle size distributions are shown at three representative magnifications (982×, 1964×, and 4910×) combined with 800C-1440M-Q-S.

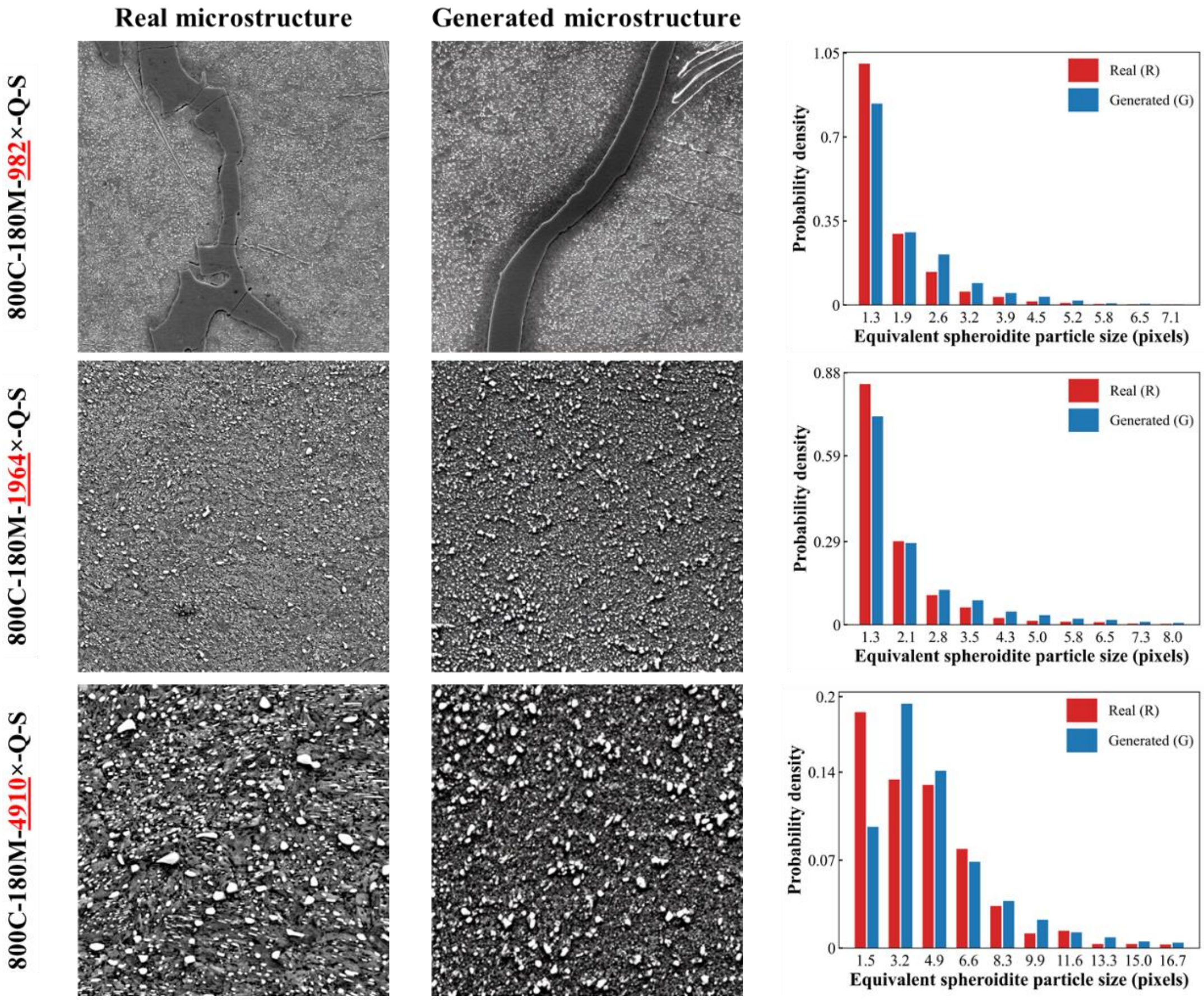

Fig. B6. Microstructure generation visualized at different magnifications under unseen conditions. Real and generated micrographs with corresponding spheroidite particle size distributions are shown at three representative magnifications (982×, 1964×, and 4910×) combined with 800C-180M-Q-S.

*c. Effect of annealing time*

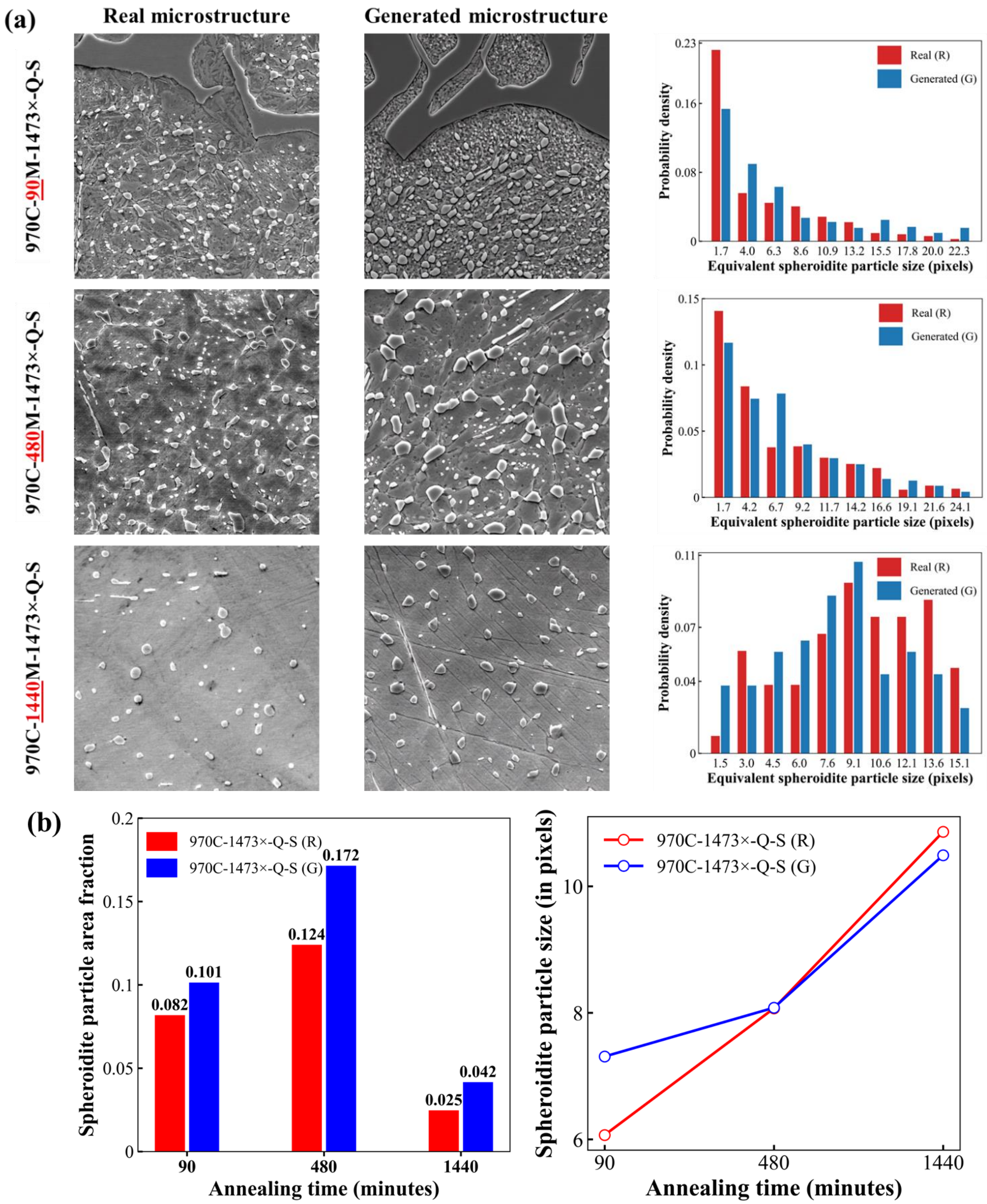

Fig. B7. Effect of annealing time on microstructure generation under seen conditions: (a) real and generated micrographs at different annealing times with corresponding spheroidite size distributions, (b) comparison of spheroidite area fraction and average particle size.

*d. Effect of cooling methods*

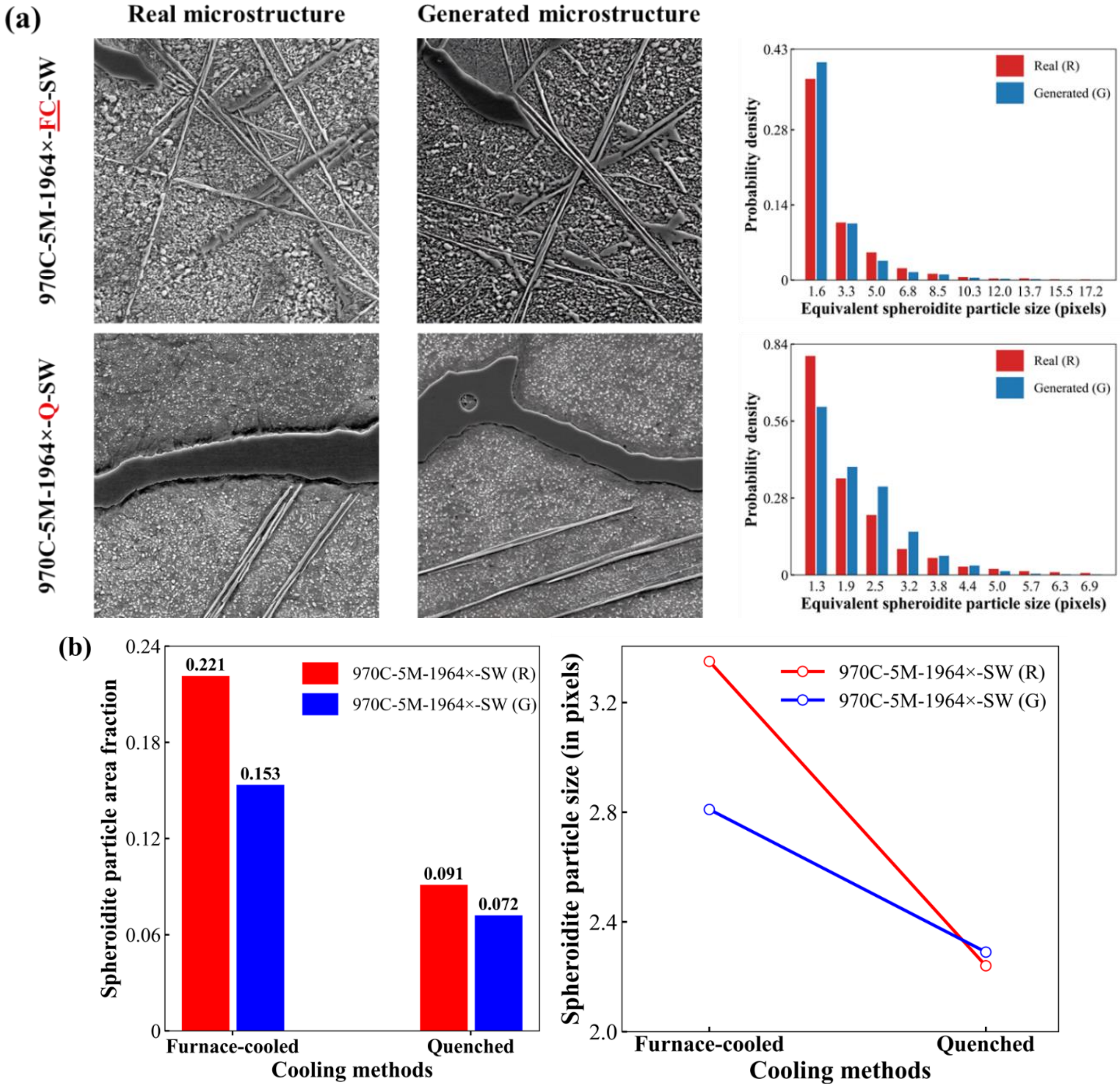

Fig. B8. Effect of cooling methods on microstructure generation under seen conditions. (a) Real and SD3.5-Large-generated micrographs with corresponding spheroidite particle size distributions, (b) spheroidite area fraction and average particle size.

B2.4. High-dimensional fidelity analysis t-SNE embedding of real and generated microstructures, SD3.5-Large (full-text) most closely reproduces the real distribution, with generated samples co-located with the real centroid and exhibiting a similar covariance footprint. SDXL performs comparably, while SD3.5-Medium remains somewhat conservative, indicating reduced spread. The parameter-only and text-only reduce bias relative to baselines but remain under-dispersed, confirming that both process variables and textual context are necessary to recover the location and diversity of the target distribution. By contrast, StyleGAN3 collapses into a compact, self-overlapping cluster, evidencing limited intra-set diversity, and DDPM exhibits a directional bias with a mis-oriented ellipse that under-covers the real support. Overall, these comparisons demonstrate that model capacity coupled with multimodal conditioning is critical for faithfully matching the real data manifold.

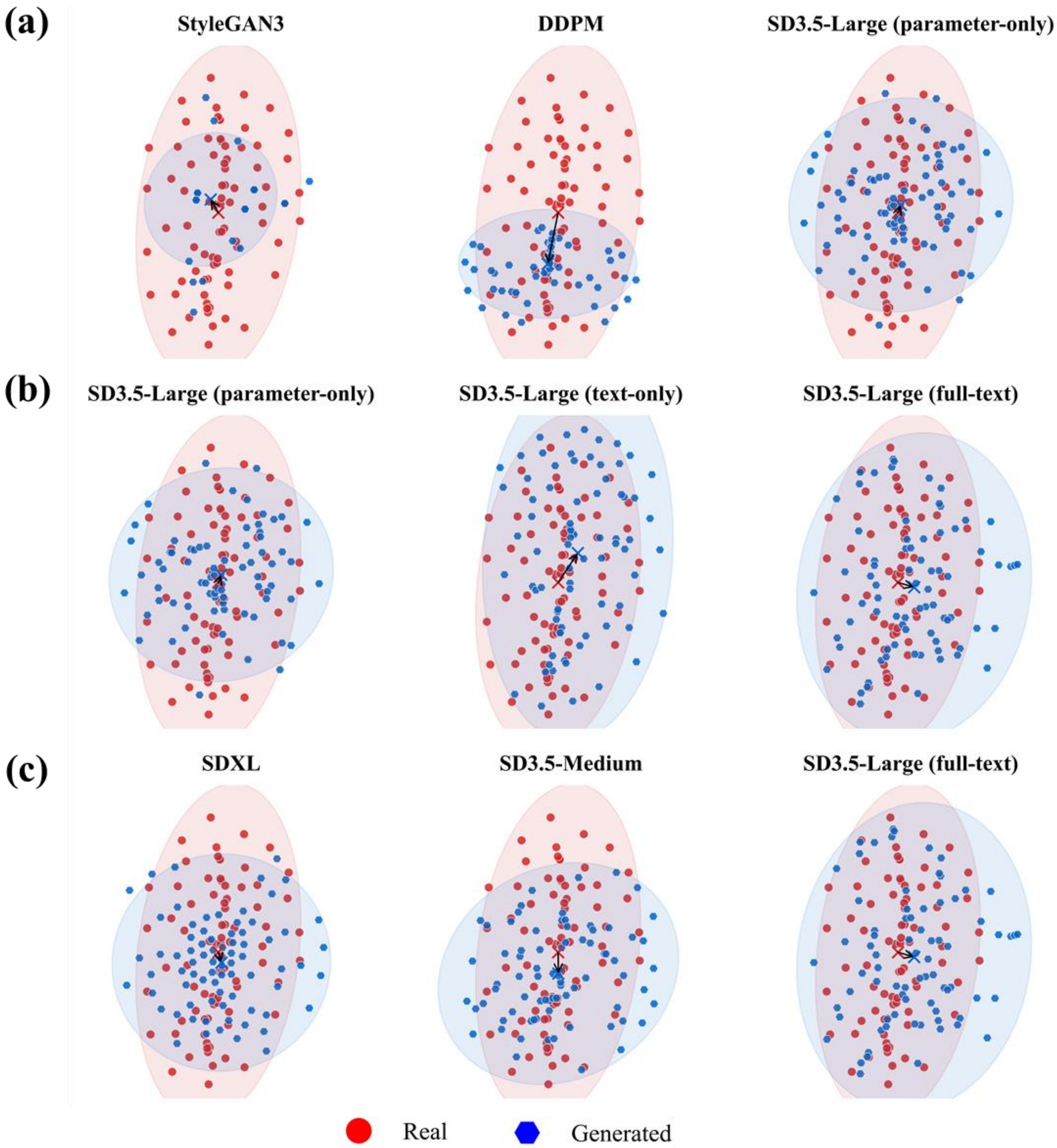

Fig. B9. t-SNE of real and generated microstructures across models and conditioning. (a) Baselines and parameter-only; (b) parameter-only, text-only, and full-text; (c) ablations: diffusion family under full-text conditioning.

The results in Table B5 demonstrate strong fidelity of the generated microstructures to real counterparts, with aspect ratio errors ranging from 4.76% to 14.67% (average 9.31%) in seen conditions and 0.70% to 13.86% (average 8.57%) in unseen scenarios, indicating robust generalization. Similarly, mean nearest neighbor distance errors average 11.54% in seen cases (5.81-22.36%) and 14.64% in unseen cases (4.79-42.12%), showing moderate to high variability, particularly in unseen conditions due to occasional outliers. Interfacial area errors remain low overall, ranging from 4.18% to12.80% (average 7.63%) in seen conditions and 0.38% to 27.49% (average 10.55% in unseen conditions, underscoring the model's effectiveness in capturing process-driven morphological features under both trained and novel parameters.

Table B5. Evaluating generative microstructures: fidelity in aspect ratios, interfacial areas, and spatial distributions across seen and unseen conditions for SD3.5-Large fine-tuned with DreamBooth and LoRA.

| Condition\descriptor | Aspect ratio ($\delta_{asp}$) | | | Mean distance of nearest neighbors (*NND*) (pixels) | | | Interfacial area (*IA*) (pixels$^{-1}$) | | |
|---|---|---|---|---|---|---|---|---|---|
| | R | G | Error (%) ↓ | R | G | Error (%) ↓ | R | G | Error (%) ↓ |
| **Seen processing** | | | | | | | | | |
| 800C-1440M-1964×-Q-S | 1.808 | 1.666 | 7.82 | 9.832 | 10.394 | 5.81 | 0.082 | 0.077 | 5.35 |
| 900C-90M-4910×-Q-S | 1.898 | 1.62 | 14.67 | 11.542 | 14.123 | 22.36 | 0.099 | 0.095 | 4.18 |
| 970C-480M-1473×-Q-S | 1.859 | 2.044 | 9.99 | 14.831 | 16.31 | 9.97 | 0.055 | 0.06 | 8.20 |
| 800C-5100M-1964×-Q-S | 1.647 | 1.569 | 4.76 | 9.035 | 9.759 | 8.02 | 0.079 | 0.069 | 12.80 |
| **Unseen processing** | | | | | | | | | |
| 800C-180M-4910×-Q-S | 1.882 | 1.686 | 10.40 | 10.378 | 11.003 | 6.02 | 0.07 | 0.068 | 3.77 |
| 800C-180M-1964×-Q-S | 1.856 | 1.869 | 0.70 | 7.235 | 7.641 | 5.62 | 0.113 | 0.144 | 27.49 |
| 970C-180M-982×-Q-S | 1.909 | 1.731 | 9.33 | 11.809 | 12.375 | 4.79 | 0.045 | 0.05 | 10.57 |
| 970C-180M-1964×-Q-S | 1.877 | 1.617 | 13.86 | 14.916 | 21.198 | 42.12 | 0.045 | 0.045 | 0.38 |

## Appendix C. Extended segmentation analysis

Table C1. Hyperparameter tuning for semantic segmentation using various pre-trained models on the UHCS subset.

| **Hyperparameters** | **VGG-16** | **VGG-19** | **ResNet-101** | **ResNet-152** |
|---|---|---|---|---|
| Encoder learning rate | 5e-5 | 8e-6 | 1e-5 | 1e-5 |
| Decoder learning rate | 5e-4 | 1e-3 | 1e-3 | 1e-3 |
| Batch size | 2 | 2 | 2 | 2 |
| Optimizer | AdamW | AdamW | AdamW | AdamW |
| Loss function | Jaccard | CCE | CCE | CCE |

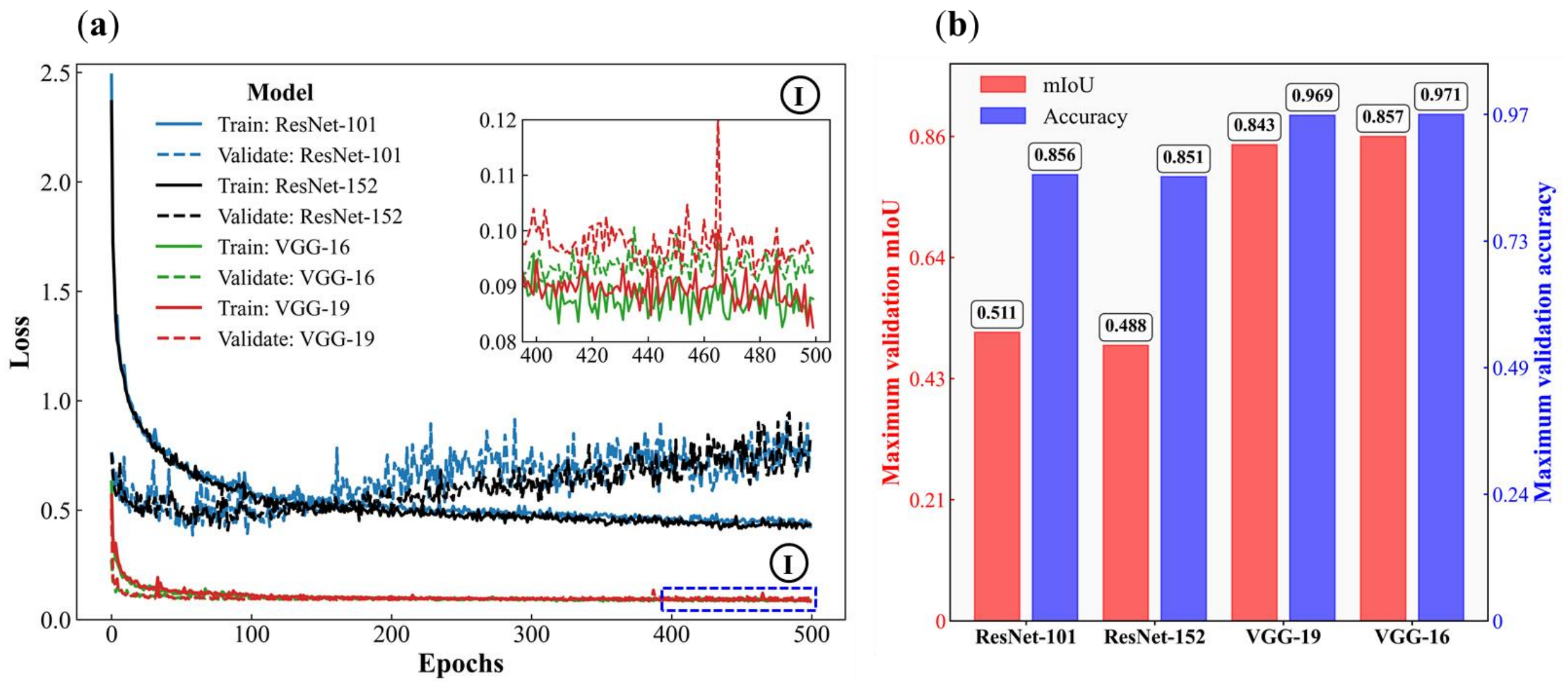

Fig. C1. Comparison of (a) losses and (b) performance metrics across several pre-trained models.

## Appendix D. Extended results on Aachen-Heerlen dataset

For generalization, we applied the same method to Aachen-Heerlen dataset [35], consisting 1705 SEM images along with annotated polygons of MA islands. We split this dataset into 1364 images for training and 341 images for validation.

Table D1. Split the Aachen-Heerlen dataset into training (1364), and validation (229 for unseen and 112 for seen during training)

| **Type (**chemical compositions of the steel**)** | **Training** | **Validation** |
|---|---|---|
| 1 | 456 | 28 |
| CFB2 | 207 | 20 |
| 5 | 162 | 18 |
| 10 | 433 | 26 |
| 11 | 218 | 20 |
| 2 | 0 | 87 |
| 3 | 0 | 3 |
| 8 | 0 | 13 |
| 12 | 0 | 126 |
| Sum | 1364 (80%) | 341 (20%) |

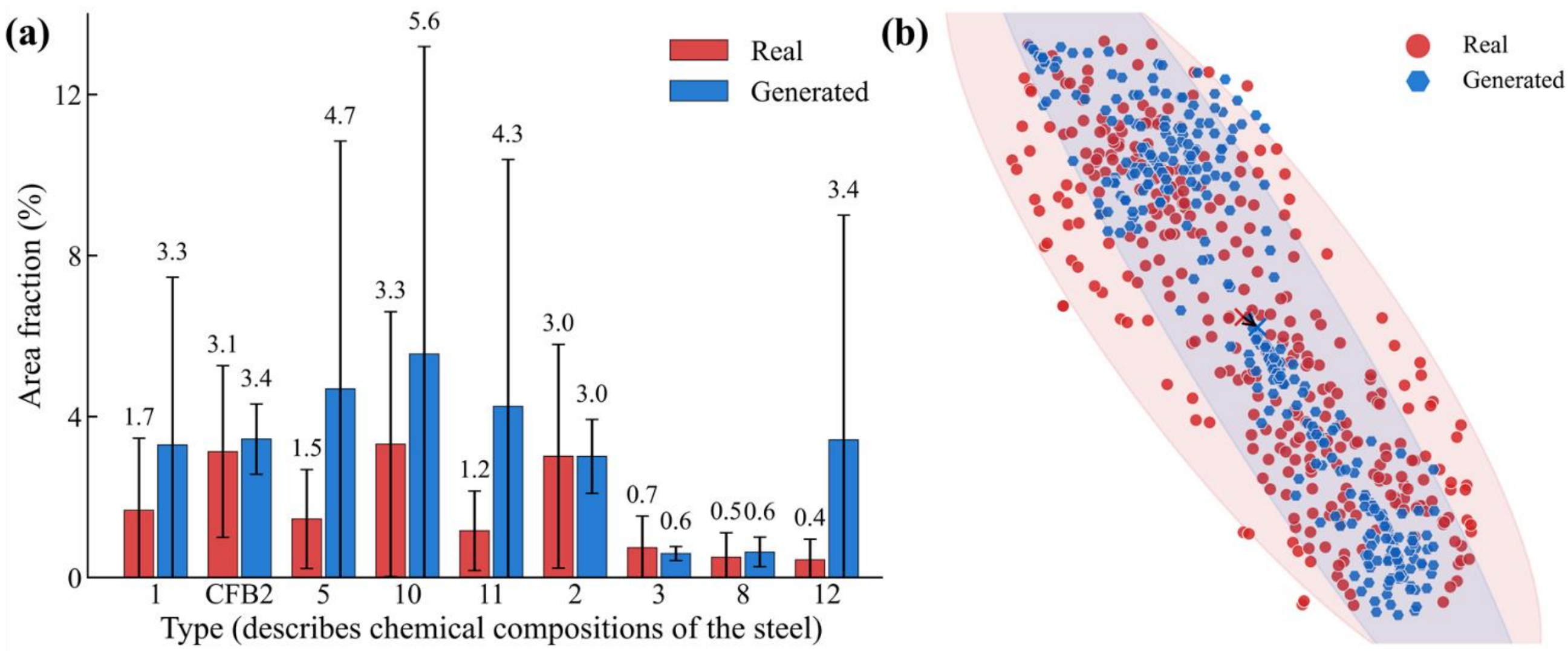

Fig. D1. Real and generated micrographs with (a) area fraction by type and (b) t-SNE overlap.

### References

[1] R. Bostanabad, Y. Zhang, X. Li, T. Kearney, L.C. Brinson, D.W. Apley, W.K. Liu, W. Chen, Computational microstructure characterization and reconstruction: Review of the state-of-the-art techniques, Prog. Mater. Sci. 95 (2018) 1–41. https://doi.org/10.1016/j.pmatsci.2018.01.005.

[2] M. Baby, A.B. Nellippallil, An information-decision framework for the multilevel co-design of products, materials, and manufacturing processes, Adv. Eng. Informatics 59 (2024) 102271. https://doi.org/10.1016/j.aei.2023.102271.

[3] M.T. Tran, H. Wang, H.W. Lee, D.K. Kim, Crystal plasticity finite element analysis of size effect on the formability of ultra-thin ferritic stainless steel sheet for fuel cell bipolar plate, Int.

J. Plast. 154 (2022) 103298. https://doi.org/10.1016/j.ijplas.2022.103298.
[4] H. Kim, K. Jung, H.W. Lee, S. Kang, D. Kim, Crystal plasticity modeling of ductile fracture locus in advanced high-strength steel, Elsevier B.V., 2025. https://doi.org/10.1016/j.rineng.2025.105720.
[5] S. Zhao, J. Song, S. Ermon, Towards Deeper Understanding of Variational Autoencoding Models, (2017). https://doi.org//10.48550/arXiv.1702.08658.
[6] B. Murgas, J. Stickel, S. Ghosh, Generative adversarial network (GAN) enabled Statistically equivalent virtual microstructures (SEVM) for modeling cold spray formed bimodal polycrystals, Npj Comput. Mater. 10 (2024) 1–14. https://doi.org/10.1038/s41524-024-01219-4.
[7] H. Thanh-Tung, T. Tran, Catastrophic forgetting and mode collapse in GANs, Proc. Int. Jt. Conf. Neural Networks (2020). https://doi.org/10.1109/IJCNN48605.2020.9207181.
[8] L. Wang, W. Chen, W. Yang, F. Bi, F.R. Yu, A State-of-the-Art Review on Image Synthesis with Generative Adversarial Networks, IEEE Access 8 (2020) 63514–63537. https://doi.org/10.1109/ACCESS.2020.2982224.
[9] J. Ho, A. Jain, P. Abbeel, Denoising diffusion probabilistic models, Adv. Neural Inf. Process. Syst. 2020-Decem (2020) 1–12.
[10] W. Zhang, G. Zhao, L. Su, Research on multi-stage topology optimization method based on latent diffusion model, Adv. Eng. Informatics 63 (2025) 102966. https://doi.org/10.1016/j.aei.2024.102966.
[11] T. Lee, H. Choi, B.J. Kim, H. Jang, D. Lee, S.W. Kim, Font conversion for steel product number recognition: A conditioned diffusion model approach, Adv. Eng. Informatics 65 (2025) 103368. https://doi.org/10.1016/j.aei.2025.103368.
[12] P. Dhariwal, A. Nichol, Diffusion Models Beat GANs on Image Synthesis, in: Adv. Neural Inf. Process. Syst., 2021: pp. 8780–8794. http://arxiv.org/abs/2105.05233.
[13] K.B. Mustapha, A survey of emerging applications of large language models for problems in mechanics, product design, and manufacturing, Adv. Eng. Informatics 64 (2025) 103066. https://doi.org/10.1016/j.aei.2024.103066.
[14] X. Kang, Z. Zhao, Optimal design of ceramic form combining stable diffusion model and GRU-Attention, Adv. Eng. Informatics 64 (2025) 103062. https://doi.org/10.1016/j.aei.2024.103062.
[15] H.C. Phan, M.T. Tran, S. Oh, D.-K. Kim, H.W. Lee, Synthetic Microstructure Generation Using Diffusion Transformers, 2 (2024) 549. https://doi.org/10.2139/ssrn.5045111.
[16] J. Tang, X. Geng, D. Li, Y. Shi, J. Tong, H. Xiao, F. Peng, Machine learning-based microstructure prediction during laser sintering of alumina, Sci. Rep. 11 (2021). https://doi.org/10.1038/s41598-021-89816-x.
[17] A. Iyer, B. Dey, A. Dasgupta, W. Chen, A. Chakraborty, A Conditional Generative Model for Predicting Material Microstructures from Processing Methods, Second Work. Mach. Learn. Phys. Sci. (NeurIPS 2019) (2019). http://arxiv.org/abs/1910.02133.
[18] M. Safiuddin, C.H. Likith Reddy, G. Vasantada, C.H.J.N.S. Harsha, S. Gangolu, Establishing Process-Structure Linkages Using Generative Adversarial Networks, in: Lect. Notes Mech. Eng., 2025: pp. 497–509. https://doi.org/10.1007/978-981-97-6367-2_39.
[19] E. Azqadan, H. Jahed, A. Arami, Predictive microstructure image generation using denoising diffusion probabilistic models, Acta Mater. 261 (2023) 119406. https://doi.org/10.1016/j.actamat.2023.119406.
[20] X. Ding, Y. Wang, Z. Xu, W.J. Welch, Z.J. Wang, Image Generation Using Continuous Conditional Generative Adversarial Networks, Intell. Syst. Ref. Libr. 217 (2022) 87–113. https://doi.org/10.1007/978-3-030-91390-8_5.
[21] X. Ding, Y. Wang, K. Zhang, Z.J. Wang, CCDM: Continuous Conditional Diffusion Models for Image Generation, 14 (2024) 1–19. http://arxiv.org/abs/2405.03546.
[22] X. Zheng, I. Watanabe, J. Paik, J. Li, X. Guo, M. Naito, Text-to-Microstructure Generation Using Generative Deep Learning, Small 2402685 (2024) 1–12.

https://doi.org/10.1002/smll.202402685.
[23] M. Heusel, H. Ramsauer, T. Unterthiner, B. Nessler, S. Hochreiter, GANs trained by a two time-scale update rule converge to a local Nash equilibrium, Adv. Neural Inf. Process. Syst. 2017-December (2017) 6627–6638. https://doi.org/10.18034/ajase.v8i1.9.
[24] P. Esser, S. Kulal, A. Blattmann, R. Entezari, J. Müller, H. Saini, Y. Levi, D. Lorenz, A. Sauer, F. Boesel, D. Podell, T. Dockhorn, Z. English, K. Lacey, A. Goodwin, Y. Marek, R. Rombach, Scaling Rectified Flow Transformers for High-Resolution Image Synthesis, Proc. 41st Int. Conf. Mach. Learn. (2024). http://arxiv.org/abs/2403.03206.
[25] N. Ruiz, Y. Li, V. Jampani, Y. Pritch, M. Rubinstein, K. Aberman, DreamBooth: Fine Tuning Text-to-Image Diffusion Models for Subject-Driven Generation, Proc. IEEE/CVF Conf. Comput. Vis. Pattern Recognit. (2023) 22500–22510. https://dreambooth.github.io/.
[26] E. Hu, Y. Shen, P. Wallis, Z. Allen-Zhu, Y. Li, S. Wang, L. Wang, W. Chen, Lora: Low-Rank Adaptation of Large Language Models, ICLR 2022 - 10th Int. Conf. Learn. Represent. (2022) 1–26.
[27] K. Simonyan, A. Zisserman, Very deep convolutional networks for large-scale image recognition, 3rd Int. Conf. Learn. Represent. ICLR 2015 - Conf. Track Proc. (2015) 1–14.
[28] B.L. DeCost, M.D. Hecht, T. Francis, B.A. Webler, Y.N. Picard, E.A. Holm, UHCSDB: UltraHigh Carbon Steel Micrograph DataBase: Tools for Exploring Large Heterogeneous Microstructure Datasets, Integr. Mater. Manuf. Innov. 6 (2017) 197–205. https://doi.org/10.1007/s40192-017-0097-0.
[29] F. Borgeaud, L., Wolf, T., Platen, P. v., Gugger, S., S., A. L. M., Wehling, W. P., Sanh, V., & Behnke, Diffusers: State-of-the-art diffusion models, Https://Github.Com/Huggingface/Diffusers (2022).
[30] M.D. Hecht, B.L. DeCost, T. Francis, E.A. Holm, Y.N. Picard, B.A. Webler, Ultrahigh Carbon Steel Micrographs, (2017). http://hdl.handle.net/11256/940.
[31] B.L. DeCost, B. Lei, T. Francis, E.A. Holm, High Throughput Quantitative Metallography for Complex Microstructures Using Deep Learning: A Case Study in Ultrahigh Carbon Steel, Microsc. Microanal. 25 (2019) 21–29. https://doi.org/10.1017/S1431927618015635.
[32] A. Radford, J. Wook, K. Chris, H. Aditya, R. Gabriel, G. Sandhini, G. Sastry, A. Askell, P. Mishkin, J. Clark, G. Krueger, I. Sutskever, Learning Transferable Visual Models From Natural Language Supervision, (2019).
[33] M. Cherti, R. Beaumont, R. Wightman, M. Wortsman, G. Ilharco, C. Gordon, C. Schuhmann, L. Schmidt, J. Jitsev, Reproducible Scaling Laws for Contrastive Language-Image Learning, 2023 IEEE/CVF Conf. Comput. Vis. Pattern Recognit. (2023) 2818–2829. https://doi.org/10.1109/cvpr52729.2023.00276.
[34] C. Raffel, N. Shazeer, A. Roberts, K. Lee, S. Narang, M. Matena, Y. Zhou, W. Li, P.J. Liu, Exploring the limits of transfer learning with a unified text-to-text transformer, J. Mach. Learn. Res. 21 (2020) 1–67.
[35] D. Iren, M. Ackermann, J. Gorfer, G. Pujar, S. Wesselmecking, U. Krupp, S. Bromuri, Aachen-Heerlen annotated steel microstructure dataset, Sci. Data 8 (2021) 1–9. https://doi.org/10.1038/s41597-021-00926-7.
[36] C. Chen, H. Ding, O. Xie, B. Yao, S. Tran, B. Sisman, Y. Xu, B. Zeng, SDXL: IMPROVING LATENT DIFFUSION MODELS FOR HIGH-RESOLUTION IMAGE SYNTHESIS, ICLR (2024) 1–22.
[37] S.Y. Liu, C.Y. Wang, H. Yin, P. Molchanov, Y.C.F. Wang, K.T. Cheng, M.H. Chen, DoRA: Weight-Decomposed Low-Rank Adaptation, Proc. Mach. Learn. Res. 235 (2024) 32100–32121.
[38] T. Karras, T. Aila, Alias-Free Generative Adversarial Networks, in: 2021: pp. 1–12. https://doi.org/35th Conference on Neural Information Processing Systems (NeurIPS 2021).
[39] A. Nichol, P. Dhariwal, A. Nichol, T. Lee, H. Choi, B.J. Kim, H. Jang, D. Lee, S.W. Kim, W. Zhang, G. Zhao, L. Su, Diffusion Models Beat GANs on Image Synthesis, Adv. Eng. Informatics 11 (2021) 8780–8794. https://doi.org/10.1016/j.aei.2025.103368.

[40] A. Nichol, P. Dhariwal, Improved Denoising Diffusion Probabilistic Models, Proc. Mach. Learn. Res. 139 (2021) 8162–8171.
[41] Y. Chen, W. Jin, M. Wang, Metallographic image segmentation of GCr15 bearing steel based on CGAN, Int. J. Appl. Electromagn. Mech. 64 (2020) 1237–1243. https://doi.org/10.3233/JAE-209441.
[42] Z. Lv, Y. Li, S. Qian, L. Wu, Semi-supervised surface defect segmentation of aluminum strips based on self-attention consistency and cross-view pseudo supervision, Adv. Eng. Informatics 66 (2025) 103389. https://doi.org/10.1016/j.aei.2025.103389.
[43] H. Li, J. Lin, Z. Liu, J. Jiao, B. Zhang, An interpretable waveform segmentation model for bearing fault diagnosis, Adv. Eng. Informatics 61 (2024) 102480. https://doi.org/10.1016/j.aei.2024.102480.
[44] J. Luengo, R. Moreno, I. Sevillano, D. Charte, A. Peláez-Vegas, M. Fernández-Moreno, P. Mesejo, F. Herrera, A tutorial on the segmentation of metallographic images: Taxonomy, new MetalDAM dataset, deep learning-based ensemble model, experimental analysis and challenges, Inf. Fusion 78 (2022) 232–253. https://doi.org/10.1016/j.inffus.2021.09.018.
[45] Q. Li, C. Ding, B. Wang, J. Jiao, W. Huang, Z. Zhu, CDARNet: A robust cross-dimensional adaptive region reconstruction network for real-time metal surface defect segmentation, Adv. Eng. Informatics 67 (2025) 103514. https://doi.org/10.1016/j.aei.2025.103514.
[46] S.Y. Kim, J.S. Kim, J.H. Lee, J.H. Kim, T.S. Han, Comparison of microstructure characterization methods by two-point correlation functions and reconstruction of 3D microstructures using 2D TEM images with high degree of phase clustering, Mater. Charact. 172 (2021) 110876. https://doi.org/10.1016/j.matchar.2021.110876.
[47] K.H. Lee, H.J. Lim, G.J. Yun, A data-driven framework for designing microstructure of multifunctional composites with deep-learned diffusion-based generative models, Eng. Appl. Artif. Intell. 129 (2024) 107590. https://doi.org/10.1016/j.engappai.2023.107590.
[48] Q. Wang, B. Wu, P. Zhu, P. Li, W. Zuo, Q. Hu, ECA-Net: Efficient channel attention for deep convolutional neural networks, Proc. IEEE Comput. Soc. Conf. Comput. Vis. Pattern Recognit. (2020) 11531–11539. https://doi.org/10.1109/CVPR42600.2020.01155.
[49] Y. Jiao, F.H. Stillinger, S. Torquato, Modeling heterogeneous materials via two-point correlation functions: Basic principles, Phys. Rev. E - Stat. Nonlinear, Soft Matter Phys. 76 (2007) 1–15. https://doi.org/10.1103/PhysRevE.76.031110.
[50] C.L.Y. Yeong, S. Torquato, Reconstructing Random Media I and II, Phys. Rev. E 58 (1998) 224–233. https://journals.aps.org/pre/pdf/10.1103/PhysRevE.57.495.
[51] M.D. Hecht, Y.N. Picard, B.A. Webler, Coarsening of Inter- and Intra-granular Proeutectoid Cementite in an Initially Pearlitic 2C-4Cr Ultrahigh Carbon Steel, Metall. Mater. Trans. A Phys. Metall. Mater. Sci. 48 (2017) 2320–2335. https://doi.org/10.1007/s11661-017-4012-2.
[52] K.H. Lee, G.J. Yun, Microstructure reconstruction using diffusion-based generative models, Mech. Adv. Mater. Struct. 0 (2023) 1–19. https://doi.org/10.1080/15376494.2023.2198528.
[53] X. Li, Y. Zhang, H. Zhao, C. Burkhart, L.C. Brinson, W. Chen, A Transfer Learning Approach for Microstructure Reconstruction and Structure-property Predictions, Sci. Rep. 8 (2018). https://doi.org/10.1038/s41598-018-31571-7.
[54] A. Radford, J.W. Kim, C. Hallacy, A. Ramesh, G. Goh, S. Agarwal, G. Sastry, A. Askell, P. Mishkin, J. Clark, G. Krueger, I. Sutskever, Learning Transferable Visual Models From Natural Language Supervision, Proc. Mach. Learn. Res. 139 (2021) 8748–8763.
[55] J. Hessel, A. Holtzman, M. Forbes, R. Le Bras, Y. Choi, CLIPScore: A Reference-free Evaluation Metric for Image Captioning, EMNLP 2021 - 2021 Conf. Empir. Methods Nat. Lang. Process. Proc. (2021) 7514–7528. https://doi.org/10.18653/v1/2021.emnlp-main.595.
[56] J. Betker, G. Goh, L. Jing, T. Brooks, J. Wang, L. Li, L. Ouyang, J. Zhuang, J. Lee, Y. Guo, W. Manassra, P. Dhariwal, C. Chu, Y. Jiao, A. Ramesh, Improving Image Generation with Better Captions, OPEN AI Publ. 2 (2023) 1.
[57] S. Jayasumana, S. Ramalingam, A. Veit, D. Glasner, A. Chakrabarti, S. Kumar, Rethinking FID: Towards a Better Evaluation Metric for Image Generation, Proc. IEEE/CVF Conf.

Comput. Vis. Pattern Recognit. (2024) 9307–9315. https://doi.org/10.1109/CVPR52733.2024.00889.
[58] T. Kynkäänniemi, T. Karras, S. Laine, J. Lehtinen, T. Aila, Improved precision and recall metric for assessing generative models, Adv. Neural Inf. Process. Syst. 32 (2019).
[59] Z. Wang, A.C. Bovik, H.R. Sheikh, E.P. Simoncelli, Image quality assessment: From error visibility to structural similarity, IEEE Trans. Image Process. 13 (2004) 600–612. https://doi.org/10.1109/TIP.2003.819861.
[60] R. Zhang, P. Isola, A.A. Efros, E. Shechtman, O. Wang, The Unreasonable Effectiveness of Deep Features as a Perceptual Metric, Proc. IEEE Comput. Soc. Conf. Comput. Vis. Pattern Recognit. (2018) 586–595. https://doi.org/10.1109/CVPR.2018.00068.
[61] W. Peebles, S. Xie, Scalable Diffusion Models with Transformers, in: 2023 IEEE/CVF Int. Conf. Comput. Vis., IEEE, 2023: pp. 4172–4182. https://doi.org/10.1109/ICCV51070.2023.00387.
[62] B.L. DeCost, T. Francis, E.A. Holm, Exploring the microstructure manifold: Image texture representations applied to ultrahigh carbon steel microstructures, Acta Mater. 133 (2017) 30–40. https://doi.org/10.1016/j.actamat.2017.05.014.